\typein[\YourChoice]{}

\def\empty{}
\ifx \YourChoice \empty
   \renewcommand{\YourChoice}{0}
\fi

\newcount\Styles
\ifnum \YourChoice=0
   \def\IncludePlot{0}
\else
   \def\IncludePlot{1}
\fi

\typein[\YourChoice]{}

\def\empty{}
\ifx \YourChoice \empty
   \renewcommand{\YourChoice}{0}
\fi

\ifnum \YourChoice=1
   \advance \Styles by 2
\fi

\ifcase \number\Styles
   \documentstyle[12pt]{article}
\or
   \documentstyle[12pt,epsf,rotate]{article}
\or
   \documentstyle[12pt,cite]{article}
\or
   \documentstyle[12pt,epsf,rotate,cite]{article}
\fi

\newcommand{\fig}[3]{
\ifnum \IncludePlot=0
   #1
   #3
\else
   #1
   #2
   #3
\fi
}

\newcommand{\newsection}[1]{\section{#1}\setcounter{equation}{0}}

\newcommand{\IsPreprint}{1}  % preprint version
\newcommand{\select}[1]{
\ifnum \IsPreprint=1
   #1
\fi
}

\begin{document}

%%%%%%%%%%%%%%%%%%%%%%%%%%%%%%%%%%%%%%%%%%%%%%%%%%%%%%%%%%%%%%%%%%%%%%%%%%%%%%%
% The titelpage
%%%%%%%%%%%%%%%%%%%%%%%%%%%%%%%%%%%%%%%%%%%%%%%%%%%%%%%%%%%%%%%%%%%%%%%%%%%%%%%

\renewcommand{\thefootnote}{\fnsymbol{footnote}}

\author{\\
\normalsize{Andrzej J. BURAS${}^{1,2}$, Markus E. LAUTENBACHER${}^{1}$,
Gaby OSTERMAIER${}^{1}$\thanks{e-mail:
{\tt {buras,lauten,gosterma}@feynman.t30.physik.tu-muenchen.de}}} \\
\ \\
{\small\sl ${}^{1}$ Physik Department, Technische Universit\"at M\"unchen,} \\
{\small\sl D-85748 Garching, Germany.}\\
{\small\sl ${}^{2}$ Max-Planck-Institut f\"ur Physik
                    -- Werner-Heisenberg-Institut,}\\
{\small\sl F\"ohringer Ring 6, D-80805 M\"unchen, Germany.}
}

\date{}

\title{
{\large\sf
\rightline{MPI-Ph/94-14}
\rightline{TUM-T31-57/94}
\rightline{March 1994}
}
\bigskip
\bigskip
{\LARGE\sf
Waiting for the Top Quark Mass, $K^+ \rightarrow \pi^+ \nu \bar\nu$,
$B_s^0$-$\bar{B}_s^0$ Mixing and CP Asymmetries in $B$-Decays}\footnote{
Supported by the German Bundesministerium f\"ur Forschung und
Technologie under contract 06 TM 732 and the CEC Science project
SC1-CT91-0729.}
}

\maketitle
\thispagestyle{empty}

\begin{abstract}
\noindent
Anticipating improved determinations of $m_t$, $\mid V_{ub}/V_{cb}
\mid$, $B_K$ and $F_B \sqrt{B_B}$ in the next five years we make an
excursion in the future in order to find a possible picture of the
unitarity triangle, of quark mixing and of CP-violation around the
year~2000. We then analyse what impact on this picture will have the
measurements of four possibly cleanest quantities:  $BR(K^+ \to \pi^+
\nu \bar\nu)$, $x_d/x_s$, $\sin(2\alpha)$ and $\sin(2\beta)$. Our
analysis shows very clearly that there is an exciting time ahead of
us.

In the course of our investigations we extend the analysis of the
unitarity triangle beyond the leading order in $\lambda$ and we derive
several useful analytic formulae for quantities of interest.
\end{abstract}

\newpage
\setcounter{page}{1}

\setcounter{footnote}{0}
\renewcommand{\thefootnote}{\arabic{footnote}}

%%%%%%%%%%%%%%%%%%%%%%%%%%%%%%%%%%%%%%%%%%%%%%%%%%%%%%%%%%%%%%%%%%%%%%%%%%%%%%%
% The main part of the paper
%%%%%%%%%%%%%%%%%%%%%%%%%%%%%%%%%%%%%%%%%%%%%%%%%%%%%%%%%%%%%%%%%%%%%%%%%%%%%%%

\newsection{Introduction}

Among the quantities studied in the rich field of rare and
CP-violating decays
\cite{winsteinwolfenstein:93,ritchiewojcicki:93,littenbergvalencia:93,burasharlander:92,nir:74,rosner:00,cassel:93}
the branching ratio $BR(K^+ \rightarrow
\pi^+ \nu \bar{\nu})$, the ratio $x_d/x_s$ of $B^o_d-\bar B^o_d$
to $B^o_s-\bar B^o_s$ mixing and a class of
CP-asymmetries in neutral B-decays, all  being essentially free from
any hadronic uncertainties, stand out as ideally suited for the
determination of the CKM parameters. Simultaneously they appear
to be in the reach of experimentalists in the next five to ten
years. The decays $K_L \rightarrow
\pi^{\circ} \nu \bar{\nu}$ and $B \rightarrow
X_s\nu \bar{\nu}$ are also theoretically very
clean but much harder to measure.

$BR(K^+ \rightarrow \pi^+ \nu \bar{\nu})$ and $x_d/x_s$ are probably the best
quantities for the determination of the CKM element
$V_{td}$ and consequently play important roles in
constraining the shape of the unitarity triangle.

The decay $K^+ \rightarrow\pi^+ \nu \bar{\nu}$ is dominated by short distance
loop diagrams involving the heavy top
quark and receives also sizable contributions from internal
charm quark exchanges. The QCD corrections to this decay have
been calculated in the leading logarithmic approximation long
time ago
\cite{ellishagelin:83,dibdunietz:91,buchallaetal:91}.
 The recent calculation
\cite{buchallaburas:94}
of next-to-leading QCD
corrections reduced consi\-derably the theoretical uncertainty
due to the choice of the renormalization scales present in the
leading order expression. Since the relevant hadronic matrix
element of the operator $\bar {s} \gamma_{\mu} (1- \gamma _{5})d~
\bar {\nu} \gamma _{\mu} (1- \gamma _{5}) \nu$ can be measured in the leading
decay $K^+ \rightarrow \pi^{\circ} e^+ \nu$, the resulting theoretical
expression for $BR(K^+ \rightarrow \pi^+ \nu \bar{\nu})$ is
only a function of the CKM parameters, the QCD scale
 $\Lambda \overline{_{MS}}$
 and the
quark masses $m_t$ and $m_c$. Moreover due to the work
of ref.~\cite{buchallaburas:94}
the scales in $m_t$ and $m_c$ are under control so that the
sensitivity of $BR(K^+ \rightarrow \pi^+ \nu \bar{\nu})$ to $m_c$ stressed in
 refs.~\cite{dib3:92,harrisrosner:92}
 is considerably
reduced. The long distance contributions to
  $K^+ \rightarrow \pi^+ \nu \bar{\nu}$ have been
considered in refs.~\cite{reinsehgal:89,hagelinlittenberg:89,luwise:94}
 and found to be very small: two to three
orders of magnitude smaller than the short distance contribution
at the level of the branching ratio.

The top quark mass dependence and the QCD corrections to $B^o-\bar B^o$
mixing cancel in the ratio $x_d/x_s$ which depends only on the CKM parameters
and SU(3)-flavour  breaking effects in the relevant hadronic matrix elements.
These SU(3) breaking effects contain much smaller theoretical uncertainties
than the hadronic matrix elements present in $x_d$ and $x_s$ separately.
The measurement of $x_d/x_s$ gives then a good determination of
the ratio $\mid V_{td}/V_{ts}\mid$ and consequently of one side of the
unitarity triangle.

The CP-asymmetry in the decay $B_d^\circ \rightarrow \psi K_S$ allows
 in the standard model
a direct measurement of the angle $\beta$ in the unitarity triangle
without any theoretical uncertainties
\cite{nir:74}. Similarly the decay
$B_d^\circ \rightarrow \pi^+ \pi^-$ gives the angle $\alpha$, although
 in this case strategies involving
other channels are necessary in order to remove hadronic
uncertainties related to penguin contributions
\cite{gronaulondon:91,gronaulondon:91a,nirquinn:91,nirquinn:91a,aleksandunietz:91}.
The determination of the angle~$\gamma$ from CP asymmetries in neutral
B-decays is more difficult but not impossible
\cite{aleksankayserlondon:93}.

At present $BR(K^+ \rightarrow \pi^+ \nu \bar{\nu})$, $x_d/x_s$ and the CP
 asymmetries in neutral B-decays
given by $\sin(2\phi_i)~(\phi_i=\alpha,\beta,\gamma)$ can be predicted using
\begin{itemize}
\item the values of $\mid V_{ub} / V_{cb}\mid$ and $\mid V_{cb}\mid$
 extracted from tree level B-decays

\item the analysis of the parameter $\epsilon _K$ describing the indirect CP
violation in K$\rightarrow \pi \pi$ decays\\
and

\item the analysis of $x_d = (\Delta M)_B/\Gamma_B$
describing the size of $B_d^\circ
-\bar{B}{_d^\circ}$ mixing
\end {itemize}

All these ingredients are subject to theoretical uncertainties
related to non-perturbative parameters entering the relevant
formulae. Moreover the last two require the value of $m_t$.
Consequently the existing predictions for
 $BR(K^+ \rightarrow \pi^+ \nu \bar{\nu}$), $x_s$ and
CP-asymmetries in B-decays are rather uncertain.

In this paper we would like to address the following questions:
\begin{itemize}
\item What accuracy of theoretical predictions for $BR(K^+
\rightarrow \pi^+ \nu \bar{\nu}$), $x_s$, $\sin(2\phi_i)$ and the
unitarity triangle could one expect around the year 2000
assuming reasonable improvements for the values of $\mid V_{cb}\mid$,
$\mid V_{ub}/V_{cb}\mid$, $m_t$ and the
non-perturbative parameters in question?

\item What would be the impact of a measurement of
 $BR(K^+ \rightarrow \pi^+ \nu \bar{\nu}$)  on the CKM
parameters and in particular on the value of $\mid V_{td}\mid$?

\item What would be the impact of a measurement of $x_s$ ?

\item What would be the impact of a measurement of $\sin(2\beta)$ and how
important would be simultaneous measurements of $\sin(2\alpha)$ and
$\sin(2\gamma)$?

\item How well should one measure $BR(K^+ \rightarrow \pi^+
\nu \bar{\nu})$, $\sin(2\beta$),
$V_{cb}$, $m_t$ and $x_d/x_s$ in order to obtain an
acceptable determination of the CKM matrix on the basis of these
five quantities alone?
\end {itemize}
As byproducts of these studies:

\begin{itemize}
\item we will update the analysis of $BR(K^+ \rightarrow \pi^+
\nu \bar{\nu}$), $x_s$, $\sin(2\phi_i)$ and of the unitarity
triangle in view of theoretical and experimental developments which
took place in 1993,

\item we will extend the analysis of the unitarity triangle beyond
the leading order in the expansion parameter $\lambda = \mid V_{us}\mid$\\
and

\item we will derive several approximate analytic formulae and
bounds which should be useful in following the developments in
this field in the 90's.
\end {itemize}

Our paper is organized as follows.
In Section 2 we extend the Wolfenstein parametrization and the
analysis of the unitarity triangle beyond the leading order in
$\lambda$ and we give improved formulae for $\sin(2\phi_i)$.
In Section 3 we collect the formulae for $\varepsilon _K$,
$B^\circ-\bar{B}^{\circ}$ mixing and
$BR(K^+ \rightarrow\pi^+ \nu \bar{\nu})$ beyond leading order in $\lambda$.
In Section 4 we list several analytic results which can be
derived using Wolfenstein parametrization beyond leading $\lambda$,
which to a very good accuracy represent exact numerical analysis.
In Section 5 we systematically address the questions posed above.
We end the paper with a brief summary and a number of conclusions.

\newsection{Cabibbo-Kobayashi-Maskawa Matrix}

\subsection{Standard Parametrization}

 We will dominantly use the standard parametrization
\cite{particle:90}
\begin{equation}\label{2.72}
V=
\left(\begin{array}{ccc}
c_{12}c_{13}&s_{12}c_{13}&s_{13}e^{-i\delta}\\ -s_{12}c_{23}
-c_{12}s_{23}s_{13}e^{i\delta}&c_{12}c_{23}-s_{12}s_{23}s_{13}e^{i\delta}&
s_{23}c_{13}\\ s_{12}s_{23}-c_{12}c_{23}s_{13}e^{i\delta}&-s_{23}c_{12}
-s_{12}c_{23}s_{13}e^{i\delta}&c_{23}c_{13}
\end{array}\right)
\end{equation}
where
 $c_{ij}=\cos\theta_{ij}$ and $s_{ij}=\sin\theta_{ij}$ with $i$ and $j$
being generation labels ($i,j=1,2,3$).
$c_{ij}$ and
$s_{ij}$ can all be chosen to be positive.
The measurements
of the CP violation in K decays force $\delta$ to be in the range
 $0<\delta<\pi$.

The extensive phenomenology of the last years
has shown that
$s_{13}$ and $s_{23}$ are small numbers: $O(10^{-3})$ and $O(10^{-2})$,
respectively. Consequently to an excellent accuracy $c_{13}=c_{23}=1$
and the four independent parameters are given as follows
\begin{equation}\label{2.73}
s_{12}=\mid V_{us}\mid, \quad s_{13}=\mid V_{ub}\mid, \quad s_{23}=\mid
V_{cb}\mid, \quad \delta,
\end{equation}
with the phase $\delta$ extracted from CP violating transitions or
loop processes sensitive to $\mid V_{td}\mid$. The latter fact is based
on the observation that
 for $0\le\delta\le\pi$, as required by the analysis of CP violation,
there is a one--to--one correspondence between $\delta$ and $|V_{td}|$
given by
\begin{equation}\label{10}
\mid V_{td}\mid=\sqrt{a^2+b^2-2 a b \cos\delta},
\qquad
a=\mid V_{cd} V_{cb}\mid,
\qquad
b=\mid V_{ud} V_{ub}\mid
\end{equation}

\subsection{Wolfenstein Parameterization Beyond Leading Order}

We will also use the Wolfenstein parametrization
\cite{wolfenstein:83}.
It is an approximate parametrization of the CKM matrix in which
each element is expanded as a power series in the small parameter
$\lambda=\mid V_{us}\mid=0.22$:
\begin{equation}\label{2.75}
V=
\left(\begin{array}{ccc}
1-{\lambda^2\over 2}&\lambda&A\lambda^3(\varrho-i\eta)\\ -\lambda&
1-{\lambda^2\over 2}&A\lambda^2\\ A\lambda^3(1-\varrho-i\eta)&-A\lambda^2&
1\end{array}\right)
+O(\lambda^4)
\end{equation}
and the set (\ref{2.73}) is replaced by
\begin{equation}\label{2.76}
\lambda, \qquad A, \qquad \varrho, \qquad \eta
\end{equation}
The Wolfenstein parameterization
has several nice features. In particular it offers in conjunction with the
unitarity triangle a very transparent geometrical
representation of the structure of the CKM matrix and allows to derive
several analytic results to be discussed below. This turns out to be very
useful in the phenomenology of rare decays and of CP violation.

When using the Wolfenstein parametrization one should remember that it
is an approximation and that in certain situations neglecting $O(\lambda^4)$
terms may give wrong results. The question then arises how to find
 $O(\lambda^4)$ and higher order terms? The point is that like in any
perturbative expansion the  $O(\lambda^4)$ and higher order terms are not
unique. This is the reason why in different papers in the literature
different  $O(\lambda^4)$ terms can be found.
The non-uniqueness of higher order terms in $\lambda$
is not troublesome however. As in any perturbation theory different choices
of expanding in $\lambda$ will result in different numerical values for the
parameters in (\ref{2.76}) extracted from the data without changing the
physics when all terms are summed up.
Here it suffices to find an expansion in $\lambda$ which allows for
simple relations between the parameters (\ref{2.73}) and (\ref{2.76}).
This will also restore the unitarity of the CKM matrix which in the
Wolfenstein parametrization as given in (\ref{2.75})
is not satisfied exactly.

To this end
we go back to (\ref{2.72}) and we impose the relations
\begin{equation}\label{2.77}
s_{12}=\lambda
\qquad
s_{23}=A \lambda^2
\qquad
s_{13} e^{-i\delta}=A \lambda^3 (\varrho-i \eta)
\end{equation}
to {\it  all orders} in $\lambda$. In view of the comments made above
this can certainly be done. It follows that
\begin{equation}\label{2.84}
\varrho=\frac{s_{13}}{s_{12}s_{23}}\cos\delta
\qquad
\eta=\frac{s_{13}}{s_{12}s_{23}}\sin\delta
\end{equation}
We observe that (\ref{2.77}) and (\ref{2.84}) represent simply
the change of variables from (\ref{2.73}) to (\ref{2.76}).
Making this change of variables in the standard parametrization
(\ref{2.72}) we find the CKM matrix as a function of
$(\lambda,A,\varrho,\eta)$ which satisfies unitarity exactly!
We also note that in view of $c_{13}=1-O(\lambda^6)$ the relations
between $s_{ij}$ and $\mid V_{ij}\mid$ in (\ref{2.73}) are
satisfied to high accuracy. The relations in (\ref{2.84}) have
been first used in ref.~\cite{schmidtlerschubert:92}.
However, our improved treatment of the unitarity
triangle presented below goes beyond the analysis of these authors.

The procedure outlined above gives automatically the corrections to the
Wolfenstein parametrization in (\ref{2.75}).  Indeed expressing
(\ref{2.72}) in terms of Wolfenstein parameters using (\ref{2.77})
and then expanding in powers of $\lambda$ we recover the
matrix in (\ref{2.75}) and in addition find explicit corrections of
$O(\lambda^4)$ and higher order terms. $V_{ub}$ remains unchanged. The
corrections to $V_{us}$ and $V_{cb}$ appear only at $O(\lambda^7)$ and
$O(\lambda^8)$, respectively.  For many practical purposes the
corrections to the real parts can also be neglected.
The essential corrections to the imaginary parts are:
\begin{equation}\label{2.83g}
\Delta V_{cd}=-iA^2 \lambda^5\eta
\qquad
\Delta V_{ts}=-iA\lambda^4\eta
\end{equation}
These two corrections have to be
taken into account in the discussion of CP violation.
On the other hand the imaginary part of $V_{cs}$ which in our expansion
in $\lambda$ appears only at $O(\lambda^6)$ can be fully neglected.

In order to improve the accuracy of the unitarity triangle discussed
below we will also include the $O(\lambda^5)$ correction to $V_{td}$
which gives
\begin{equation}\label{2.83d}
 V_{td}= A\lambda^3 (1-\bar\varrho-i\bar\eta)
\end{equation}
with
\begin{equation}\label{2.88d}
\bar\varrho=\varrho (1-\frac{\lambda^2}{2})
\qquad
\bar\eta=\eta (1-\frac{\lambda^2}{2}).
\end{equation}
In order to derive analytic results we need accurate explicit expressions
for $\lambda_i=V_{td}V_{ts}^*$ where $i=c,t$. We have
\begin{equation}\label{2.51}
 Im\lambda_t= -Im\lambda_c=\eta A^2\lambda^5
\end{equation}
\begin{equation}\label{2.52}
 Re\lambda_c=-\lambda (1-\frac{\lambda^2}{2})
\end{equation}
\begin{equation}\label{2.53}
 Re\lambda_t= -(1-\frac{\lambda^2}{2}) A^2\lambda^5 (1-\bar\varrho)
\end{equation}
Expressions (\ref{2.51}) and (\ref{2.52}) represent to an accuracy of
0.2\% the exact formulae obtained using (\ref{2.72}). The expression
(\ref{2.53}) deviates by at most 2\% from the exact formula in the
full range of parameters considered.
In order to keep the analytic
expressions in sections 3 and 4 in a transparent form
we have dropped a small $O(\lambda^7)$ term in deriving (\ref{2.53}).
After inserting the expressions (\ref{2.51})--(\ref{2.53}) in exact
formulae for quantities of interest, further expansion in $\lambda$
should not be made.

\subsection{Unitarity Triangle Beyond Leading Order}
The unitarity of the CKM-matrix provides us with several relations
of which
\begin{equation}\label{2.87h}
V_{ud}V_{ub}^* + V_{cd}V_{cb}^* + V_{td}V_{tb}^* =0
\end{equation}
is the most useful one.
In the complex plane the relation (\ref{2.87h}) can be represented as
a triangle, the so-called ``unitarity--triangle'' (UT).
Phenomenologically this triangle is very interesting as it involves
simultaneously the elements $V_{ub}$, $V_{cb}$ and $V_{td}$ which are
under extensive discussion at present.

In the usual analyses of the unitarity triangle only terms $O(\lambda^3)$
are kept in (\ref{2.87h})
\cite{burasharlander:92,nir:74,harrisrosner:92,schmidtlerschubert:92,dibdunietzgilman:90,alilondon:93}.
It is however straightforward to include the
next-to-leading $O(\lambda^5)$ terms.
We note first that
\begin{equation}\label{2.88a}
V_{cd}V_{cb}^*=-A\lambda^3+O(\lambda^7).
\end{equation}
Thus to an excellent accuracy $V_{cd}V_{cb}^*$ is real with
$\mid V_{cd}V_{cb}^*\mid=A\lambda^3$.
Keeping $O(\lambda^5)$ corrections and rescaling all terms in (\ref{2.87h})
by $A \lambda^3$
we find
\begin{equation}\label{2.88b}
 \frac{1}{A\lambda^3}V_{ud}V_{ub}^*
=\bar\varrho+i\bar\eta
\qquad,
\qquad
 \frac{1}{A\lambda^3}V_{td}V_{tb}^*
=1-(\bar\varrho+i\bar\eta)
\end{equation}
with $\bar\varrho$ and $\bar\eta$ defined in (\ref{2.88d}).
Thus we can represent (\ref{2.87h}) as the unitarity triangle
in the complex $(\bar\varrho,\bar\eta)$
plane. This is  shown in fig.~\ref{fig:triangle}.
 The length of the side $CB$ which lies
on the real axis equals unity when eq.~(\ref{2.87h}) is rescaled by
$V_{cd}V_{cb}^*$. We observe that beyond the leading order
in $\lambda$ the point A {\it does not} correspond to  $(\varrho,\eta)$ but to
 $(\bar\varrho,\bar\eta)$.
Clearly within 3\% accuracy $\bar\varrho=\varrho$ and $\bar\eta=\eta$.
Yet in the distant future the accuracy of experimental results and
theoretical calculations may improve considerably so that the more
accurate formulation given here will be appropriate. For instance
the experiments at LHC should measure $ \sin(2\beta) $ to an accuracy
of 2--3\% \cite{camilleri:93}.

\fig{
\begin{figure}[htb]
}{
\vspace{0.05in}
\centerline{
% adjust actual size of fig. in text here
\epsfysize=2in
\epsffile{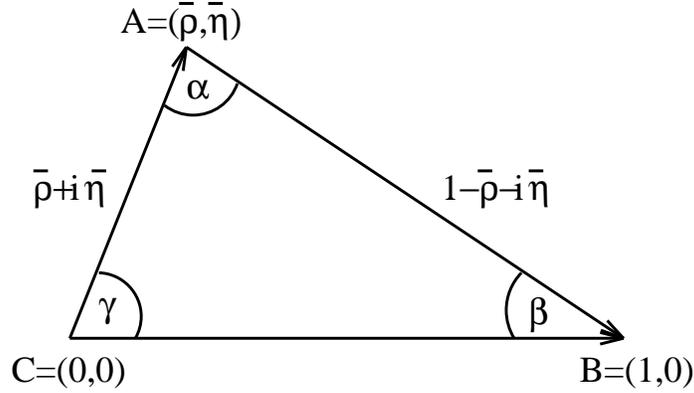}
% use this if rotation of fig. is needed
% \rotate[r]{
% \epsffile{triangle.ps}
% }
}
\vspace{0.05in}
}{
\caption[]{\small\sl
Unitarity triangle in the complex $(\bar\varrho,\bar\eta)$ plane.
\label{fig:triangle}}
\end{figure}
}

Using simple trigonometry one can calculate
$\sin(2\phi_i$) in terms of $(\bar\varrho,\bar\eta)$
with the result:
\begin{equation}\label{2.89}
\sin(2\alpha)=\frac{2\bar\eta(\bar\eta^2+\bar\varrho^2-\bar\varrho)}
  {(\bar\varrho^2+\bar\eta^2)((1-\bar\varrho)^2
  +\bar\eta^2)}
\end{equation}
\begin{equation}\label{2.90}
\sin(2\beta)=\frac{2\bar\eta(1-\bar\varrho)}{(1-\bar\varrho)^2 + \bar\eta^2}
\end{equation}
 \begin{equation}\label{2.91}
\sin(2\gamma)=\frac{2\bar\varrho\bar\eta}{\bar\varrho^2+\bar\eta^2}=
\frac{2\varrho\eta}{\varrho^2+\eta^2}
\end{equation}
The lengths $CA$ and $BA$ in the
rescaled triangle of fig.\ 1 to be denoted by $R_b$ and $R_t$,
respectively, are given by
\begin{equation}\label{2.94}
R_b \equiv \frac{\mid V_{ud}V^*_{ub}\mid}{\mid V_{cd}V^*_{cb}\mid}
= \sqrt{\bar\varrho^2 +\bar\eta^2}
= (1-\frac{\lambda^2}{2})\frac{1}{\lambda}
\left| \frac{V_{ub}}{V_{cb}} \right|
\end{equation}
\begin{equation}\label{2.95}
R_t \equiv \frac{\mid V_{td}V^*_{tb}\mid}{\mid V_{cd}V^*_{cb}\mid} =
 \sqrt{(1-\bar\varrho)^2 +\bar\eta^2}
=\frac{1}{\lambda} \left| \frac{V_{td}}{V_{cb}} \right|
\end{equation}
The expressions for $R_b$ and $R_t$ given here in terms of
$(\bar\varrho, \bar\eta)$
are excellent approximations. Clearly $R_b$ and $R_t$
can also be determined by measuring two of the angles $\phi_i$:
\begin{equation}\label{2.96}
R_b=\frac{\sin(\beta)}{\sin(\alpha)}=
\frac{\sin(\alpha+\gamma)}{\sin(\alpha)}=
\frac{\sin(\beta)}{\sin(\gamma+\beta)}
\end{equation}
\begin{equation}\label{2.97}
R_t=\frac{\sin(\gamma)}{\sin(\alpha)}=
\frac{\sin(\alpha+\beta)}{\sin(\alpha)}=
\frac{\sin(\gamma)}{\sin(\gamma+\beta)}
\end{equation}

\newsection{Basic Formulae}

\subsection{Constraint from $\varepsilon_{K}$}

The usual box diagram calculation together with the experimental value
for $\varepsilon_K$ specifies a hyperbola in the $(\varrho, \eta)$
plane with $\eta > 0$ \cite{harrisrosner:92,dibdunietzgilman:90}. With
our new coordinates $(\bar\varrho,\bar\eta)$ we get
\begin{equation}\label{100}
\bar\eta \left[(1-\bar\varrho) A^2 \eta_2 S(x_t)
+ P_0(\varepsilon) \right] A^2 B_K = 0.223
\end{equation}
Here

\begin{equation}\label{102}
P_0(\varepsilon) = \left[ \eta_3 S(x_c,x_t) - \eta_1 x_c \right]
\frac{1}{\lambda^4}
\end{equation}

\begin{equation}\label{103}
S(x_c,x_t) = x_c \left[ \ln \frac{x_t}{x_c} - \frac{3x_t}{4(1-x_t)}
\left(1+ \frac{x_t}{1-x_t}\ln x_t\right)\right]
\end{equation}

\begin{equation}\label{104}
S(x_t) = x_t \left[ \frac{1}{4} + \frac{9}{4} \frac {1}{(1-x_t)} - \frac
{3}{2} \frac{1}{(1-x_t)^2}\right] + \frac{3}{2}
\left[ \frac{x_t}{x_t-1}\right]^3 \ln x_t
\end{equation}
where $x_i = m^2_i/M^2_W$. $B_K$
is the renormalization group
invariant
non-perturbative parameter describing the size of
$<\bar{K^0} \mid (\bar s d)_{V-A}(\bar s d)_{V-A}\mid K^0>$
and $\eta_i$ represent QCD corrections to the box diagrams.

In our numerical analysis we will use
\begin{equation}\label{105}
\eta_1 = 1.1 \mbox{\cite{herrlichnierste:93}},\qquad \eta_2 = 0.57
\mbox{\cite{burasjaminweisz:90}},\qquad
\eta_3=0.36 \mbox{\cite{kaufmanetal:88,buchallaetal:90,dattaetal:90,flynn:90}}
\mbox{(leading order)}
\end{equation}
The values for $B_K$ are specified below.

\subsection{$B^o-\bar B^o$ Mixing}

The experimental knowledge of the $B^o_d-\bar B^o_d$ mixing described by
the parameter $x_d = \Delta M/\Gamma_B$ determines $\mid V_{td} \mid$.
Using the usual formulae for box diagrams with top quark exchanges one
finds
\begin{equation}\label{105a}
x_d = \mid V_{td} \mid ^2 P(B^o_d-\bar B^o_d) S(x_t)
\end{equation}
where
\begin{equation}\label{105b}
P(B^o_d-\bar B^o_d) = 3.89 \cdot 10^3 \left[ \frac{\tau_{B_d}}{1.5\,ps}
\right] \left[ \frac{F_{B_d} \sqrt{B_{B_d}}}{200 ~MeV} \right] ^2
\left[ \frac{\eta_B}{0.55} \right]
\end{equation}
and consequently
\begin{equation}\label{106}
\mid V_{td} \mid = A \lambda ^3 R_t ,\qquad
 R_t = 1.63 \cdot \frac{R_0}{\sqrt{ S(x_t)}}.
\end{equation}
Here
\begin{equation}\label{107}
R_0 \equiv \sqrt{ \frac{x_d}{0.72}} \left[ \frac{200 MeV}{F_{B_d}
 \sqrt{B_{B_d}}}
\right] \left[ \frac{0.038}{\kappa} \right] \sqrt{ \frac{0.55}{\eta_B}}
\end{equation}
and
\begin{equation}\label{107a}
\kappa \equiv \mid V_{cb} \mid \left[\frac{\tau_B}{1.5\,ps}\right]^{0.5}
\end{equation}
with $\tau_B$ being the B-meson life-time.
$\eta_B$ is the QCD factor analogous to $\eta_2$ and calculated
to be $\eta_B = 0.55$ \cite{burasjaminweisz:90}.
$F_{B_d}$ is the B-meson decay constant and $B_{B_d}$
denotes a non-perturbative
parameter analogous to $B_K$. The values of $x_d$, $F_{B_d} \sqrt{ B_{B_d}}$
and $|V_{cb}|$  will be specified below.

It is well known (see for instance \cite{alilondon:93}) that the
accuracy of the determination of $\mid V_{td} \mid$ and $R_t$ can be
considerably  improved by measuring simultaneously the $B^o_s-\bar
B^o_s$ mixing described by $x_s$. Defining the ratio
\begin{equation}\label{107b}
R_{ds} = \frac{\tau_{B_d}}{\tau_{B_s}} \cdot \frac{m_{B_d}}{m_{B_s}}
\left[ \frac{F_{B_d} \sqrt{B_{B_d}}}{F_{B_s} \sqrt{B_{B_s}}} \right]^2
\end{equation}
we find
\begin{equation}\label{107c}
R_t = \frac{1}{\sqrt{R_{ds}}} \sqrt{\frac{x_d}{x_s}} \frac{1}{\lambda}
\sqrt{1-\lambda^2(1-2 \varrho)}
\end{equation}
and using (\ref{106}) the matrix element $\mid V_{td} \mid$. The last
factor in (\ref{107c}) describes small departure of $\mid V_{ts} \mid$
from $\mid V_{cb} \mid$. The $\varrho$ dependence in (\ref{107c})
can safely be
neglected. In this way $R_t$ does not depend neither on $m_t$ nor on $
\mid V_{cb} \mid$. Since it is easier to  calculate $R_{ds}$  than
$R_0$, formula (\ref{107c}) gives a much more reliable determination
of $R_t$ than (\ref{106}) provided $x_s$ has been measured.

\subsection{The Rare Decay $K^{+} \to \pi^{+} \nu \bar\nu$}

The $K^{+} \to \pi^{+} \nu \bar\nu$ branching ratio for one single
neutrino flavor $l~(l = e,\mu,\tau)$ is given by
\begin{equation}\label{107d}
BR(K^{+} \to \pi^{+} \nu \bar\nu) =
 \frac{\alpha^2 BR(K^{+} \to \pi^0 e^+ \nu)}{V_{us}^2 2 \pi^2
\sin^4\theta_W} \cdot \mid V_{cs}^\ast V_{cd} X_{NL}^l +
V_{ts}^\ast V_{td} X(x_t) \mid^2
\end{equation}
Summing over three neutrino flavors,  using eqs.~(\ref{2.51})--(\ref{2.53})
and setting
\begin{equation}\label{107e}
\alpha = \frac{1}{128} \qquad \sin ^2 \theta_W = 0.23 \qquad
 BR(K^{+} \to \pi^0 e^+ \nu) = 4.82 \cdot 10^{-2}
\end{equation}
we obtain
\begin{equation}\label{108}
BR(K^{+} \to \pi^{+} \nu \bar\nu) = 4.64 \cdot 10^{-11} A^4 X^2(x_t)
\frac{1}{\sigma} \left[ (\sigma \bar\eta)^2 +
\frac{2}{3} \left(\varrho^e_0 - \bar\varrho \right)^2 +
\frac{1}{3} \left(\varrho^{\tau}_0 - \bar\varrho \right)^2 \right]
\end{equation}
with
\begin{equation}\label{109}
\varrho^l_0 = 1 + \frac{P^l_0}{A^2 X(x_t)} \qquad
P^l_0 = \frac{X^l_{NL}}{\lambda^4} \qquad
\sigma = \left( \frac{1}{1- \frac{\lambda^2}{2}} \right)^2
\end{equation}
The function $X(x_t)$ is given as follow
\begin{eqnarray}\label{109a}
X(x_t) &  = & \eta_{\rm X} \cdot X_0(x_t) \\
X_0(x_t) & =& \frac{x}{8} \left[ - \frac{2+x}{1-x} + \frac{3x
-6}{(1-x)^2} \ln x \right] \qquad {\rm with} \quad \eta_X = 0.985
\end{eqnarray}
where $\eta_{\rm X}$ is the NLO correction calculated in
ref.~\cite{buchallaburas:93b} .
For determining $P^l_0$ given in tab.~\ref{tab:plopar} we take the
NLO results for $X^l_{NL}$ of ref.~\cite{buchallaburas:94}. Here
$m_c \equiv \overline{m}_c(m_c)$.

\begin{table}[htb]
\caption[]{\small\sl
	Values of $P_0^l$ for various $\Lambda_{\overline {\rm
	MS}}~[MeV]$ and $m_c~[GeV]$}
\vspace{0.05in}
\begin{center}
\begin{tabular}{|c||c|c|c||c|c|c|}
\hline
\multicolumn{1}{|c||}{  } & \multicolumn{3}{c||}{$P_0^e $} &
\multicolumn{3}{c|}{$P_0^\tau $} \\
\hline
$\Lambda_{\rm{\overline{MS}}} \backslash m_c$ & 1.25 & 1.30 &
      1.35 & 1.25 &1.30 & 1.35 \\
\hline
\hline
 0.20 &  0.457 &  0.494 &  0.531 &  0.312 &  0.342 &  0.373 \\
\hline
 0.25 &  0.441 &  0.477 &  0.515 &  0.296 &  0.326 &  0.357 \\
\hline
 0.30 &  0.425 &  0.461 &  0.498 &  0.280 &  0.309 &  0.340 \\
\hline
 0.35 &  0.408 &  0.444 &  0.480 &  0.262 &  0.292 &  0.322 \\
\hline
\end{tabular}
\end{center}
\label{tab:plopar}
\end{table}

The measured value of BR($K^{+} \to \pi^{+} \nu \bar\nu$)
determines  an ellipse in the $(\bar\varrho,\bar\eta)$ plane  centered at
$(\varrho_0,0)$ with
\begin{equation}\label{110}
\varrho_0 = 1 + \frac{\bar{P_0}(K^+)}{A^2 X(x_t)} \qquad
\bar{P_0}(K^+) = \frac{2}{3} P^e_0 + \frac{1}{3} P^\tau_0
\end{equation}
and having the axes squared
\begin{equation}\label{110a}
\bar\varrho_1^2 = r^2_0 \qquad \bar\eta_1^2 = \left( \frac{r_0}{\sigma}
\right)^2
\end{equation}
where
\begin{equation}\label{111}
r^2_0 = \frac{1}{A^4 X^2(x_t)} \left[
\frac{\sigma \cdot BR(K^{+} \to \pi^{+} \nu \bar\nu)}{4.64 \cdot 10^{-11}}
- \frac{2}{9} \left( P^e_0 - P^\tau_0 \right)^2 \right].
\end{equation}
The last term in (\ref{111}) is very small and can safely be
neglected.

The ellipse defined by $r_0$, $\varrho_0$ and $\sigma$ given above
intersects for the allowed range of parameters
with the circle (\ref{2.94}) . This allows to determine $\bar\varrho$ and
$\bar\eta$  with
\begin{equation}\label{113}
\bar\varrho = \frac{1}{1-\sigma^2} \left( \varrho_0 - \sqrt{\varrho_0^2
-(1-\sigma^2)(\varrho_0^2 -r_0^2+\sigma R_b^2)} \right) \qquad
\bar\eta = \sqrt{R_b^2 -\bar\varrho^2}
\end{equation}
and consequently
\begin{equation}\label{113aa}
R_t^2 = 1+R_b^2 - 2 \bar\varrho
\end{equation}
where $\bar\eta$ is assumed to be positive.

In the leading order of the Wolfenstein parametrization
\begin{equation}\label{113ab}
\sigma \to 1 \qquad \bar\eta \to \eta \qquad \bar\varrho \to \varrho
\end{equation}
and $BR(K^+ \to \pi^+ \nu \bar\nu)$ determines a circle in the
$(\varrho,\eta)$ plane centered at $(\varrho_0,0)$ and having the radius
$r_0$ of (\ref{111}) with $\sigma =1$. Formulae (\ref{113}) and
(\ref{113aa}) simplify then to
\begin{equation}\label{113a}
R_t^2 = 1 + R_b^2 + \frac{r_0^2 - R_b^2}{\varrho_0} - \varrho_0 \qquad
\varrho = \frac{1}{2} \left( \varrho_0 + \frac{R_b^2 - r_0^2}{\varrho_0}
\right)
\end{equation}
in accordance with ref.~\cite{buchallaburas:94}.

\subsection{$B^o$-Decays and Superweak Models}

Although the CP-asymmetries in $B^0-$decays in which the final state is a CP
eigenstate offer a way to measure the angles of the unitarity triangle,
they may in principle
fail to distinguish the standard model from superweak models.
 As discussed by G\'erard and Nakada \cite{gerardnakada:91}
and
by Liu and Wolfenstein \cite{liuwolfenstein:87}, non-vanishing asymmetries are
also expected in superweak scenarios. In order to rule out superweak models
one has to measure the asymmetries in two distinct channels and find that
they differ from each other. As an example consider
 $B^0\to \psi K_S$ $(CP=-1)$ and $B^0 \to \pi^+\pi^-$ $(CP=1)$
for which the time integrated asymmetries are
\begin{equation}\label{113c}
 A_{CP}(\psi K_S)=-\sin(2\beta) \frac{x_d}{1+x_d^2},~~~~~~~~~~
   A_{CP}(\pi^+\pi^-)=-\sin(2\alpha) \frac{x_d}{1+x_d^2}
\end{equation}
Generally these two
asymmetries could differ in the standard model both in sign and
magnitude. In a superweak model however these asymmetries differ
only by the sign of the CP-parity of the final state. Yet as emphasized
by Winstein \cite{winstein:91} if $\sin 2\beta =-\sin 2\alpha$ it will be
impossible to distinguish the standard model result from superweak
models. This will happen for any $\bar\varrho >0$ and $\bar\eta$ given
by \cite{winstein:91}
\begin{equation}\label{113d}
 \bar\eta=(1-\bar\varrho) \sqrt{\frac{\bar\varrho}{2-\bar\varrho}}
\end{equation}
as can be easily verified using (\ref{2.89}) and (\ref{2.90}).
Consequently $(\bar\varrho,\bar\eta)$ must lie sufficiently away from
the curve of eq.~(\ref{113d}) in order to rule out the superweak
scenario on the basis of $B^0$-decays to CP eigenstates.  We will
investigate in section \ref{sec:pheno} whether this is likely to happen
in the future experiments.

\newsection{Analytic Results}

Now, we want to give a list of results following from the formulae above
which can be presented in an analytic form. Some of these results appeared
already in the literature.

\subsection{Lower Bounds on $m_t$ and $B_K$ from $\varepsilon_K$}

The hyperbola (\ref{100}) intersects the circle given by (\ref{2.94}) in two
points. It is usually stated in the literature that one of these
points corresponds to $\bar\varrho < 0$ and the other one to
$\bar\varrho > 0$. For
most values of $A$, $B_K$ and $m_t$ this is in fact true. However, with
decreasing $A$, $B_K$ and $m_t$, the hyperbola (\ref{100}) moves away
from the origin of the $(\bar\varrho, \bar\eta)$ plane and both solutions can
appear for $\bar\varrho < 0$. For sufficiently low values of these parameters
the hyperbola and the circle only touch each other at a small negative
value of $\bar\varrho$. In this way a lower bound for $m_t$ as a function of
$B_K$, $V_{cb}$ and $\mid V_{ub}/V_{cb} \mid$ can be found.

With an accurate approximation for $S(x_t)$
\begin{equation}\label{114}
S(x_t) = 0.784 \cdot x_t^{0.76}
\end{equation}
one can derive an analytic lower bound on $m_t$ \cite{buras:93}, which
to an accuracy of 2\% reproduces the exact numerical result. It is
given by

\begin{equation}\label{115}
(m_t)_{min} = M_W \left[ \frac{1}{2 A^2} \left(\frac{1}{A^2 B_K R_b}
 - 1.2 \right) \right]^{0.658}
\end{equation}

A detailed analysis of (\ref{115}) can be found in
ref.~\cite{buras:93}.
 Here we want
to stress that once $m_t$ has been determined, the same analysis gives
the minimal value of $B_K$ consistent with measured $\varepsilon_K$ as
a function of $\mid V_{cb} \mid$ and $\mid V_{ub} / V_{cb} \mid$. We
find
\begin{equation}\label{116}
(B_K)_{min} = \left[ A^2 R_b \left( 2 x_t^{0.76} A^2 + 1.2 \right)
\right]^{-1}
\end{equation}
Choosing $m_t = 180~GeV$ we show $(B_K)_{min}$ as a function of $\mid
V_{cb} \mid$ for different values of $\mid V_{ub} / V_{cb} \mid$ in
fig.~\ref{fig:bkmin}.  For
lower values of $m_t$ the bound is stronger. We observe that for $m_t
\leq 180~GeV$, $\mid V_{ub}/V_{cb} \mid \leq 0.10$ and
$\mid V_{cb} \mid \leq 0.040$ only values $B_K > 0.55$ are consistent
with $\varepsilon_{K}$ in the framework of the standard model.
\fig{
\begin{figure}[htb]
}{
\vspace{0.05in}
\centerline{
% adjust actual size of fig. in text here
\epsfysize=5in
%\epsffile{bkmin.ps}
% use this if rotation of fig. is needed
 \rotate[r]{
 \epsffile{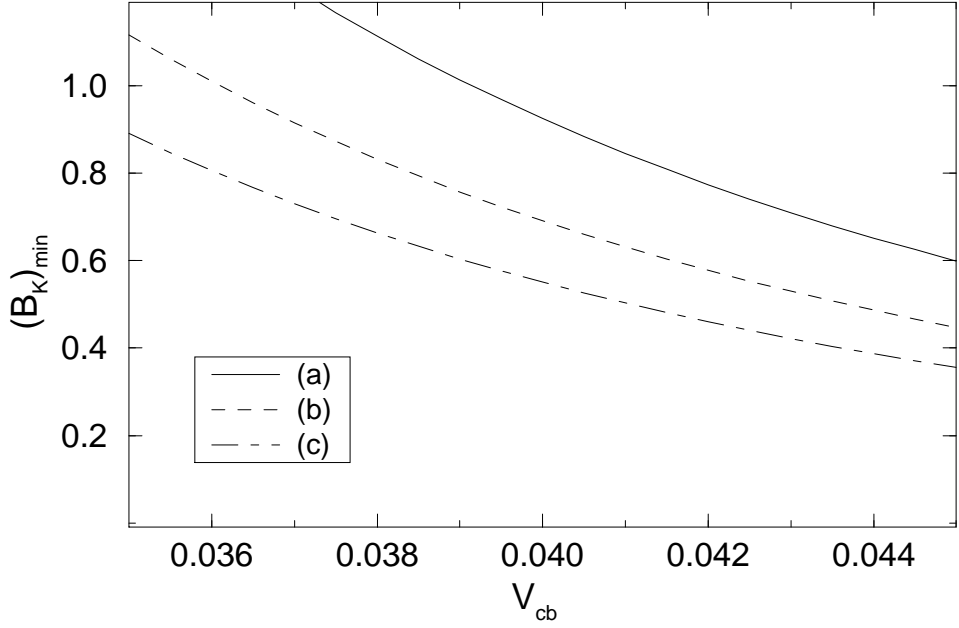}
 }
}
\vspace{0.05in}
}{
\caption[]{\small\sl
Lower bound on $B_K$ for $\mid V_{ub}/V_{cb} \mid = 0.06$
(a), $\mid V_{ub}/V_{cb} \mid = 0.08$ (b), $\mid V_{ub}/
V_{cb} \mid = 0.10$ (c) from $\varepsilon_K$ and $m_t < 180\,GeV$.
\label{fig:bkmin}}
\end{figure}
}

\subsection{Upper Bound on $\sin(2 \beta)$}

For the present range of $R_b$ the angle $\beta$ is smaller than $45^o$.
This allows to derive an upper bound on $\sin(2 \beta)$, which depends
only on $R_b$. As shown in  fig.~\ref{fig:betamax} it is found to be
\begin{equation}\label{117}
(\sin(2 \beta))_{max} = 2 R_b \sqrt{1- R_b^2}
\end{equation}
\fig{
\begin{figure}[htb]
}{
\vspace{0.05in}
\centerline{
% adjust actual size of fig. in text here
\epsfysize=2in
\epsffile{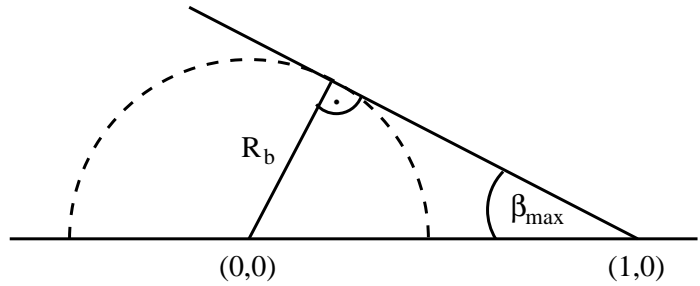}
% use this if rotation of fig. is needed
% \rotate[r]{
% \epsffile{betamax.ps}
% }
}
\vspace{0.00in}
}{
\caption[]{
\small\sl Determination of $(\sin(2\beta))_{max}$.
\label{fig:betamax}}
\end{figure}
}

This implies
\begin{equation}\label{118}
(\sin(2 \beta))_{max} = \left\{ \begin{array}{r}
0.795 \quad (\beta_{max} = 26.3^o) \qquad \mid V_{ub}/V_{cb}\mid~ =~ 0.10\\
0.663 \quad (\beta_{max} = 20.8^o) \qquad \mid V_{ub}/V_{cb}\mid~ =~ 0.08\\
0.513 \quad (\beta_{max} = 15.4^o) \qquad \mid V_{ub}/V_{cb}\mid~ =~ 0.06
\end{array}\right.
\end{equation}
A lower bound on $\sin(2 \beta)$ can only be found numerically as it
depends on $\bar\eta$. The result can be inferred from
our numerical analysis in section~\ref{sec:pheno}.

\subsection{$\sin(2 \beta)$ from $\varepsilon_K$ and $B^o-\bar B^o$
Mixing}

Combining (\ref{100}) and (\ref{106}) one can derive an analytic
formula for $\sin(2 \beta)$. We find
\begin{equation}\label{119}
\sin(2 \beta) = \frac{1}{1.33 \cdot A^2 \eta_2 R_0^2} \left[
\frac{0.223}{A^2 B_K} - \bar\eta P_0(\varepsilon) \right].
\end{equation}
$ P_0(\varepsilon)$ is weakly dependent on
$m_t$ and for $150 \leq m_t \leq 180 ~GeV$ one has $P_0(\varepsilon) \approx
0.26 \pm 0.02$. As $\bar\eta \leq 0.45$ for $ \mid V_{ub}/V_{cb} \mid
\leq 0.1$ the first term in parenthesis is generally by a factor of 2--3
larger then the second term. Since this dominant term is independent
of $m_t$, the values for $\sin(2 \beta)$ extracted from $\varepsilon_K$
and $B^o-\bar B^o$ mixing show only a weak dependence on $m_t$ as
stressed in particular in ref.~\cite{rosner:00}.

\subsection{Ambiguity in $\bar\varrho$}

It is well known that in the analysis of $\varepsilon_K$ with fixed $
\mid V_{ub}/V_{cb} \mid$  and $V_{cb}$ one gets two solutions for
$(\bar\varrho,\bar\eta)$ with $\bar\eta$ being larger for the solution with
larger
$\bar\varrho$ . The solution of this ambiguity in $\bar\varrho$ is very
important for
CP-violating decays $K_L^0 \to \pi^0 e^+ e^-$, $K_L^0 \to \pi^0 \nu
\bar\nu$ and the CP-asymmetries in B-decays governed by $\sin(2
\beta)$, because $BR(K_L^0 \to \pi^0 e^+ e^-)$, $BR(K_L^0 \to \pi^0 \nu
\bar\nu)$ and $\sin(2 \beta)$ are larger for the solution with
larger $\bar\varrho$. The
preferred solution in  searches of CP violation corresponds in most
cases to $\bar\varrho \geq 0$.

This should be contrasted with any CP conserving transition sensitive
to $\mid V_{td} \mid$, such as $B^o-\bar B^o$ mixing ,
$K^+ \to \pi^+ \nu \bar\nu$, $K_L \to \mu \bar\mu$, $B \to \mu \bar\mu$
which for  given values of $m_t$, $F_B
\sqrt{B_B}$ , $V_{cb}$, $x_d$ determine
uniquely the value of $\bar\varrho$.

Although several analysis of this determination have been presented in
the literature (see in particular ref.~\cite{harrisrosner:92}) , we
think it is useful to have simple analytic expressions helping to
answer immediately, whether the favored solution $\bar\varrho \geq 0$
is chosen.

\subsubsection{$B^o-\bar B^o$ Mixing}

We require that $R_t \leq \sqrt{1+R_b^2}$. Then for a given value of
$R_b$ one gets
a positive $\bar\varrho$. Using the analytic formula
(\ref{114}) and introducing the ``scaling'' variable
\cite{burasharlander:92}
\begin{equation}\label{120}
z(B^o_d) = m_t \left[\frac{\kappa}{0.038} \right]^{1.32}
\end{equation}
we find using (\ref{106}) and (\ref{107}) the condition
\begin{equation}\label{121}
F_{B_d} \sqrt{B_{B_d}} \geq \sqrt\frac{0.55}{\eta_B} \sqrt\frac{x_d}{0.72}
\left[ \frac{179 ~GeV}{z(B^o_d)} \right]^{0.76} \cdot
\frac{200 ~MeV}{\sqrt{1+R_b^2}}
\end{equation}

When this inequality is satisfied the favored solution with
$\bar\varrho \geq 0$ is bound to be chosen. Setting $\eta_B = 0.55$ we
plot in fig.~\ref{fig:fbmax} the smallest value of $F_B \sqrt{B_B}$
consistent with (\ref{121}) as a function of $z(B^o_d)$ for different
values of $|V_{ub}/V_{cb}|$ and $x_d = 0.72$. We observe that for
$z(B^o_d) \leq 180~GeV$ one needs $F_{B_d} \sqrt{B_{B_d}} \geq 180~MeV$
in order to have $\bar\varrho \geq 0$.

Using (\ref{107c}) we can also find  a minimal value for $x_s$
consistent with $R_t \leq \sqrt{1+R_b^2}$. One gets
to a very good approximation
\begin{equation}\label{121a}
(x_s)_{min} = \frac{x_d}{R_{ds} \lambda^2} \cdot
\frac{1}{\sqrt{1+R_b^2}}
\end{equation}
For $R_{ds} = 1$ and $R_b = 1/3$ we have $(x_s)_{min}
\simeq  18.6 \cdot x_d$.

\fig{
\begin{figure}[htb]
}{
\vspace{0.05in}
\centerline{
% adjust actual size of fig. in text here
\epsfysize=5in
%\epsffile{fbmax.ps}
% use this if rotation of fig. is needed
 \rotate[r]{
 \epsffile{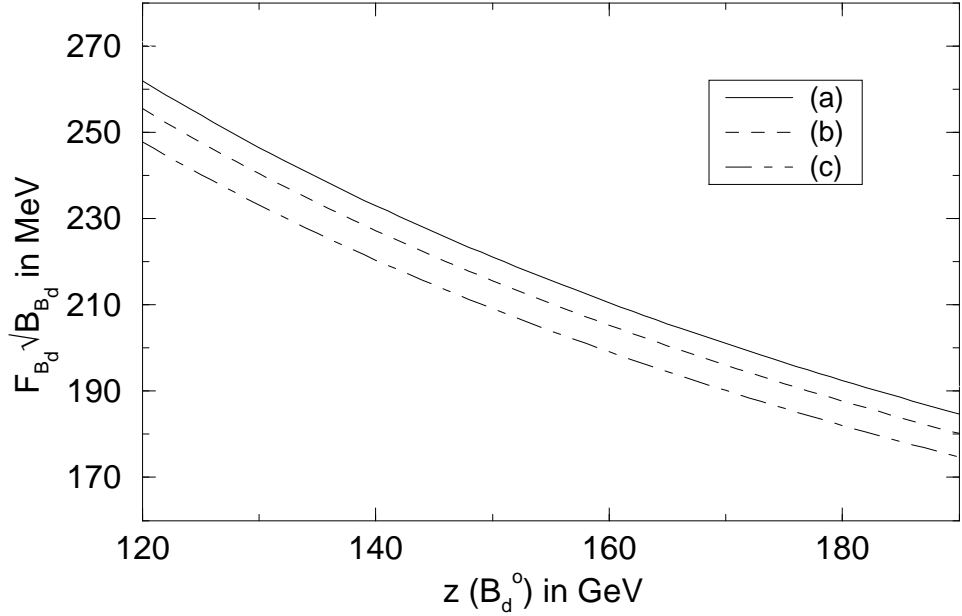}
 }
}
\vspace{0.02in}
}{
\caption[]{\small\sl
Lower bound on $F_{B_d} \sqrt{B_{B_d}}$  for  $\mid V_{ub}/V_{cb} \mid = 0.06$
(a), $\mid V_{ub}/V_{cb} \mid = 0.08$ (b), $\mid V_{ub}/
V_{cb} \mid = 0.10$ (c) necessary for $\bar\varrho \ge 0$.
\label{fig:fbmax}}
\end{figure}
}

\subsubsection{$K^+ \to \pi^+ \nu \bar\nu$}

An analogous condition can be derived from the decay
$K^+ \to \pi^+ \nu \bar\nu$
by requiring $\sqrt{{\varrho_0}^2 + (R_b \sigma)^2} \geq r_0$ with
$\varrho_0$ and $r_0$ defined in
(\ref{110}) and (\ref{111}), respectively. Neglecting the tiny
contribution of the second term in (\ref{111}), using the formula
\begin{equation}\label{122}
X(x_t) = 0.65 \cdot x_t^{0.575}
\end{equation}
which reproduces the function $X(x_t)$ to an accuracy of
0.5\% for the range of $m_t$ considered in this paper
and introducing the variable \cite{burasharlander:92}
\begin{equation}\label{123}
z(K^+) =m_t \cdot \left[ \frac{V_{cb}}{0.038}
\right]^{1.74}
\end{equation}
we find the condition
\begin{eqnarray}\label{124}
BR(K^+ \to \pi^+ \nu \bar\nu) \leq 4.64 \cdot 10^{-11}
\frac{1}{\sigma} &
 \left\{ \left[0.40 \left( \frac{z(K^+)}{M_W} \right)^{1.15}
           + \bar P_0(K^+) \right]^2  \right.  \\
    &  \left. + 0.16 \left(\frac{z(K^+)}{M_W} \right)^{2.30}
    (R_b\,\sigma)^2 \right\} \nonumber
\end{eqnarray}
This bound is shown in fig.~\ref{fig:brkmax} as a function of the variable
$z(K^+)$. Although this solution is
welcome in searches for CP violation, the experimental bound on $BR(K^+
\to \pi^+ \nu \bar\nu)$,  which could be reached in the coming years
\cite{kuno:92} ,
will be most probably above it.
\fig{
\begin{figure}[htb]
}{
\vspace{0.05in}
\centerline{
% adjust actual size of fig. in text here
\epsfysize=5in
%\epsffile{brkmax.ps}
% use this if rotation of fig. is needed
 \rotate[r]{
 \epsffile{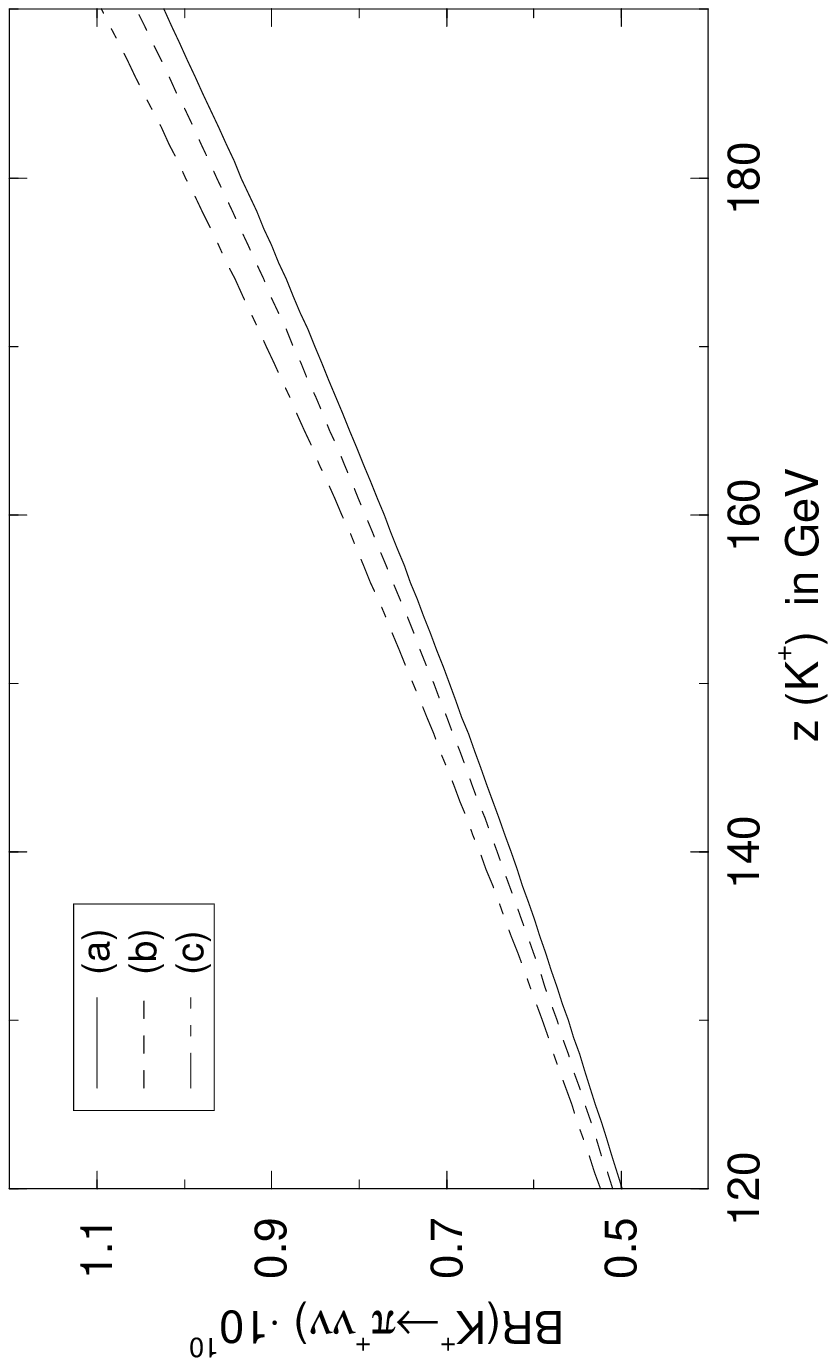}
 }
}
\vspace{0.03in}
}{
\caption[]{\small\sl
Upper bound on $BR(K^+ \to \pi^+ \nu \bar\nu)$ for
  $\mid V_{ub}/V_{cb} \mid = 0.06$
(a), $\mid V_{ub}/V_{cb} \mid = 0.08$ (b), $\mid V_{ub}/
V_{cb} \mid = 0.10$ (c) necessary for $\bar\varrho \ge 0$.
\label{fig:brkmax}}
\end{figure}
}

\newsection{Phenomenological Analysis}\label{sec:pheno}

\subsection{First Look}
\label{subsec:first}

In order to  describe the situation of 1994 after a possible top quark
discovery we make first the following
choices for the relevant parameters:

\bigskip
\underline{Range I}

\begin{equation}\label{200}
\begin{array}{rclrcl}
\left| V_{cb} \right| & = &  0.038 \pm 0.004 &
\mid V_{ub}/V_{cb} \mid & = & 0.08 \pm 0.02 \\
B_K & = & 0.7 \pm 0.2 & \sqrt{B_{B_d}} F_{B_d} & = & (200 \pm 30)~MeV \\
x_d & = & 0.72 \pm 0.08 & m_t & = & (165 \pm 15)~GeV \\
\end{array}
\end{equation}

The values of $\mid V_{cb} \mid$ and $\mid V_{ub}/V_{cb} \mid$ given
here are consistent with the recent summary in \cite{stone:93}.
 The values of $B_K$ cover comfortably the range of most recent lattice
$(B_K = 0.825 \pm 0.027)$ \cite{kilcupetal:93} and 1/N $(B_K = 0.7 \pm
0.1)$ \cite{bardeenetal:88} results. They also touch the range of
values obtained in the hadron duality approach $(B_K = 0.4 \pm 0.1)$
\cite{pradesetal:91}.  $\sqrt{B_{B_d}} F_{B_d} $ given here is in the
ball park of various lattice and QCD sum rule estimates
\cite{kronfeldmackenzie:93}. $x_d$ is in accordance with the most
recent average of CLEO and ARGUS data \cite{cassel:93} and it is
compatible with the LEP data. We set $\tau_B = 1.5~ps$ \cite{lueth:93}
in the whole analysis because the existing small error on $\tau_B$
$(\Delta \tau_B = \pm 0.04~ps)$ has only a very small impact on our
numerical results.

The choice for $m_t$ requires certainly an explanation. The high
precision electroweak studies give in the standard model typically $m_t
\simeq 165 \pm 30~GeV$ where the central value corresponds to $m_H =
300~GeV$ \cite{hollik:93}. Since we work in the standard model we
expect that $m_t$ will be found in this range. A top quark discovery at
TEVATRON will certainly narrow this range by at least a factor of two.
It is of interest to see what impact this would have for the
phenomenology considered here. At this level of accuracy one has to
state how $m_t$ is defined. The QCD corrections to $\varepsilon_K$,
$B^o-\bar B^o$ mixing and $K^+ \to \pi^+ \nu \bar\nu$ used here
correspond to the running top quark mass in the $\overline{MS}$ scheme
evaluated at $m_t$ i.e. $m_t$ in (\ref{200}) and in all formulae of
this paper stands for $\overline{m_t} (m_t)$. The physical top quark
mass as the pole of the renormalized propagator is then given by
\begin{equation}\label{201}
m_t^{phys}(m_t) = m_t \left[ 1+ \frac{4 \alpha_s (m_t)}{3 \pi} \right]
\end{equation}
For the range of $m_t$ considered here $m_t^{phys}$ is by $7 \pm
1~GeV$ higher than $m_t$.

For $\Lambda_{\overline{MS}}$ and $m_c$ affecting $BR(K^+ \to \pi^+ \nu
\bar\nu)$ we use
\begin{equation}\label{Lmsmcrange1}
\Lambda_{\overline{MS}} = (0.275 \pm\,0.075)\,GeV
\qquad
m_c \equiv \overline{m}_c(m_c) = (1.3 \pm 0.05)\,GeV
\end{equation}

\select{
\medskip
\hrule
\smallskip
\noindent
{\bf Note of caution to the reader of the preprint version:} \\
Due due limitations of our plot program the labels in
figs.~\ref{fig:utriag}, \ref{fig:sinesab}, \ref{fig:sinesbc} and
\ref{fig:sin2bbr} read $\varrho$ and $\eta$ but in fact should read
$\bar\varrho$ and $\bar\eta$.
\smallskip
\hrule
\medskip
}

In fig.~\ref{fig:utriag}\,(I) we show the resulting unitarity
triangle . To this end the analysis of $\varepsilon_K$ and
of $B_d^o-\bar{B_d^o}$~mixing have been used. In
tab.~\ref{tab:rangefirst} we show the resulting ranges for $\delta$,
$\sin(2\phi_i)$, $BR(K^+ \to \pi^+ \nu \bar\nu)$, $ \mid V_{td} \mid$
and $x_s$ corresponding to the choice of the parameters in
(\ref{200}). In calculating $x_s$ we have set $R_{ds} = 1$.

We observe:
\begin{itemize}
\item The uncertainty in the value of $\sin(2\beta)$ is moderate.
	We find $\sin(2\beta) \simeq 0.59 \pm 0.21$.
        Consequently a large asym\-me\-try $A_{CP}(\psi K_s)$ is
	ex\-pected.
      In par\-tic\-ular $\sin(2\beta)~\geq~0.38$.
\item The uncertainties in $\sin(2\alpha)$ and in
	$\sin(2\gamma)$ are huge.
\item Similarly the uncertainties in the predicted values of
	$BR(K^+ \to \pi^+ \nu \bar\nu)$, $ \mid V_{td} \mid$ and $x_s$
	are large
\end{itemize}
\begin{table}[htb]
\caption[]{\small\sl
 Ranges for scan of basic parameters for range I as of eq.~(\ref{200}).
 split according to the two different solutions for the CKM
 phase $\delta$ in the first and second quadrant
 }
\vspace{0.02in}
\begin{center}
\begin{tabular}{|c||c|c||c|c|}
\hline
\multicolumn{1}{|c||}{  } & \multicolumn{2}{c||}{1. Quadrant} &
\multicolumn{2}{c|}{2. Quadrant} \\
\hline
    & Min & Max & Min & Max \\
\hline
\hline
$\delta$  & 44.5 & 90.0 & 90.0 & 135.9  \\
\hline
$\sin(2 \alpha)$        &  -0.67 &   0.74 &   0.50 &   1.00  \\
\hline
$\sin(2 \beta)$     &   0.50 &   0.80 &   0.38 &   0.74  \\
\hline
$\sin(2 \gamma)$         &   0 &   1.00 &  -1.00 &   0  \\
\hline
$\mid V_{td} \mid\cdot 10^3$    &  6.9 & 10.0 &  8.6 & 11.8  \\
\hline
$x_s$		            &  10.8 &  24.2 &   7.7 &  14.4  \\
\hline
$BR(K^+ \to \pi^+ \nu \bar\nu)  \cdot 10^{10}$ &
  0.62 &   1.39 &   0.67 &   1.46  \\
\hline
\end{tabular}
\end{center}
\label{tab:rangefirst}
\end{table}
\fig{
\begin{figure}[htb]
}{
\vspace{0.05in}
\centerline{
% adjust actual size of fig. in text here
\epsfysize=3in
%\epsffile{utriag.ps}
% use this if rotation of fig. is needed
 \rotate[r]{
 \epsffile{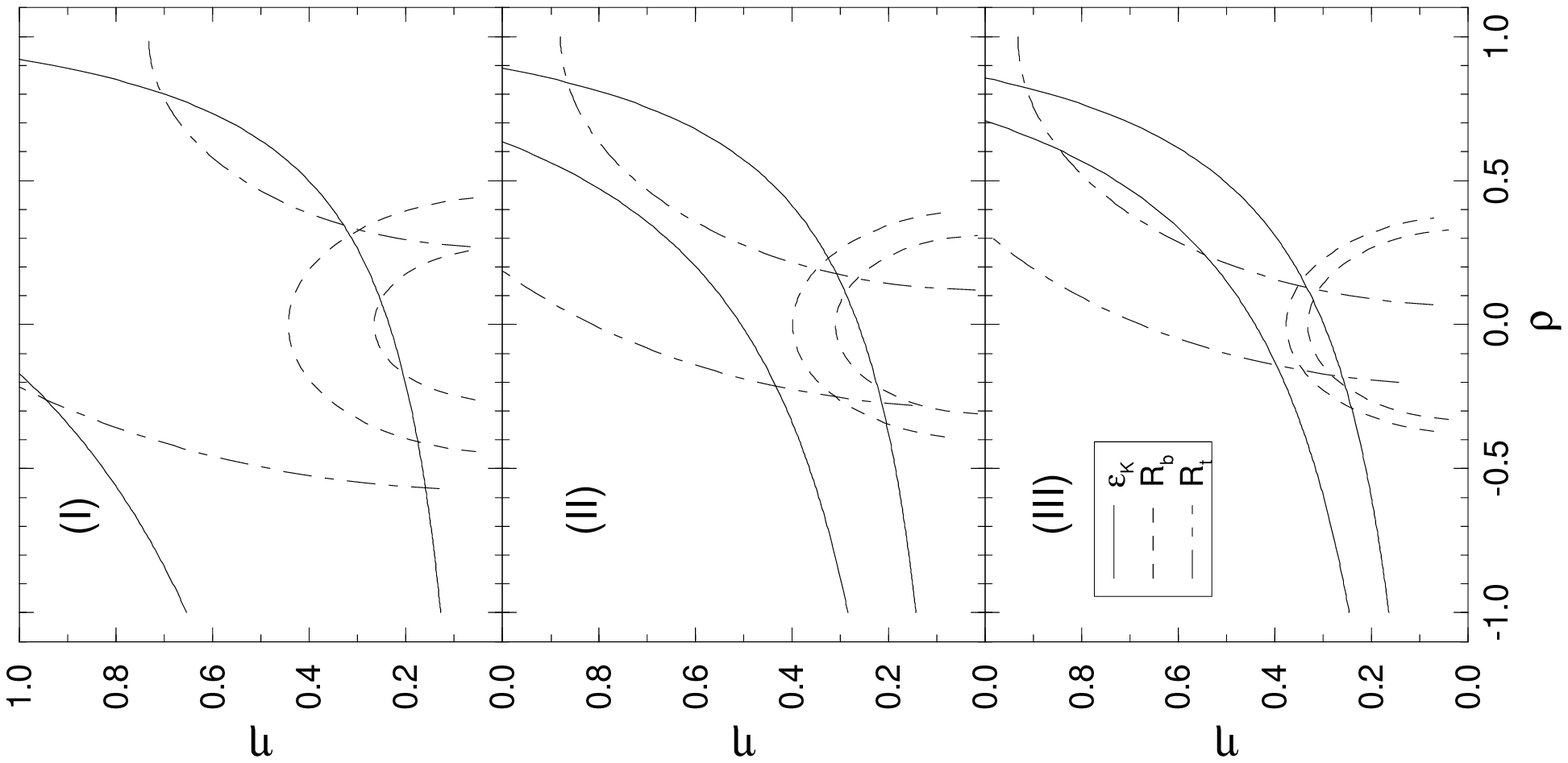}
 }
}
\vspace{0.05in}
}{
\caption[]{\small\sl
Unitarity triangle in the $(\bar\varrho,\bar\eta)$
plane determined by $\varepsilon_{K}$, $\mid V_{ub}/V_{cb} \mid$ and $x_d$
using ranges (I)--(III) as of eqs.~(\ref{200}), (\ref{210}) and
(\ref{211}), respectively.
\label{fig:utriag}}
\end{figure}
}

\subsection{A Look in the Future}
\label{subsec:look}

It is to be expected that the uncertainties in (\ref{200}) will be
reduced in the next five years through the improved determinations of
$\mid V_{cb} \mid$ and $\mid V_{ub}/V_{cb} \mid$ at CLEO II
\cite{cassel:93}, the
improved measurements of $x_d$ and the discovery of top. We also
anticipate that the extensive efforts of theorists, in particular
using the lattice methods, will considerably reduce the errors
on $B_K$ and $\sqrt{B_B} F_B$.

We consider the following ranges of parameters:

\bigskip
\underline{Range II}
\begin{equation}\label{210}
\begin{array}{rclrcl}
\left| V_{cb} \right| & = &  0.040 \pm 0.002 &
\mid V_{ub}/V_{cb} \mid& = & 0.08 \pm 0.01 \\
B_K & = & 0.75 \pm 0.07 & \sqrt{B_{B_d}} F_{B_d} & = & (185 \pm 15)~MeV \\
x_d & = & 0.72 \pm 0.04 & m_t & = & (170 \pm 7)~GeV \\
\end{array}
\end{equation}

\underline{Range III}
\begin{equation}\label{211}
\begin{array}{rclrcl}
\left| V_{cb} \right| & = &  0.040 \pm 0.001 &
\mid V_{ub}/V_{cb} \mid & = & 0.08 \pm 0.005 \\
B_K & = & 0.75 \pm 0.05 & \sqrt{B_{B_d}} F_{B_d} & = & (185 \pm 10)~MeV \\
x_d & = & 0.72 \pm 0.04 & m_t & = & (170 \pm 5)~GeV \\
\end{array}
\end{equation}

For $\Lambda_{\overline{MS}}$ and $m_c$ we use
\begin{equation}\label{Lmsmcrange2}
\Lambda_{\overline{MS}} = 0.3\,GeV
\qquad
m_c = 1.3\,GeV
\end{equation}

For each range we repeat the analysis of subsection
\ref{subsec:first}. The results are given in fig.~\ref{fig:utriag}\,(II)
and (III) and tabs.~\ref{tab:rangeb} and \ref{tab:rangec}.

\begin{table}[htb]
\caption[]{
	\small\sl
	Same as in tab.~\ref{tab:rangefirst} but for range~II as of
      eq.~(\ref{210}).
}
\vspace{0.05in}
\begin{center}
\begin{tabular}{|c||c|c||c|c|}
\hline
\multicolumn{1}{|c||}{  } & \multicolumn{2}{c||}{1. Quadrant} &
\multicolumn{2}{c|}{2. Quadrant} \\
\hline
    & Min & Max & Min & Max \\
\hline
\hline
$\delta$  & 60.9 & 90.0 & 90.0 & 122.5 \\
\hline
$\sin(2 \alpha)$                   &  -0.30 &   0.69 &   0.57 &   1.00  \\
\hline
$\sin(2 \beta)$ 	      &   0.57 &   0.73 &   0.46 &   0.69  \\
\hline
$\sin(2 \gamma)$     	      &   0 &   0.85 &  -0.91 &  0  \\
\hline
$\mid V_{td} \mid\cdot 10^3$      &  8.1  &  9.8 &  9.0  & 10.8  \\
\hline
$x_s$		    &  11.2 &  17.6 &   9.1 &  13.0  \\
\hline
$BR(K^+ \to \pi^+ \nu \bar\nu)  \cdot 10^{10}$
&   0.83 &   1.22 &   0.86 &   1.3  \\
\hline
\end{tabular}
\end{center}
\label{tab:rangeb}
\end{table}

\begin{table}[htb]
\caption[]{
	\small\sl
	Same as in tab.~\ref{tab:rangefirst} but for range~III as of
      eq.~(\ref{211}).
}
\vspace{0.05in}
\begin{center}
\begin{tabular}{|c||c|c||c|c|}
\hline
\multicolumn{1}{|c||}{  } & \multicolumn{2}{c||}{1. Quadrant} &
\multicolumn{2}{c|}{2. Quadrant} \\
\hline
    & Min & Max & Min & Max \\
\hline
\hline
$\delta$ & 69.0 & 90.0 & 90.0 & 113.7 \\
\hline
$\sin(2 \alpha)$ & 0.01 & 0.66 & 0.60 & 0.99 \\
\hline
$\sin(2 \beta)$ & 0.60 & 0.70 & 0.52 & 0.66 \\
\hline
$\sin(2 \gamma)$     & 0 & 0.67 & -0.69 & 0 \\
\hline
$\mid V_{td} \mid\cdot 10^3$ & 8.4 & 9.6 & 9.1 & 10.4 \\
\hline
$x_s$  & 11.9 & 15.6& 10.1 & 13.3 	\\
\hline
$BR(K^+ \to \pi^+ \nu \bar\nu)  \cdot 10^{10}$ &
0.88 & 1.12 & 0.92 & 1.18 \\
\hline
\end{tabular}
\end{center}
\label{tab:rangec}
\end{table}

We observe:
\begin{itemize}
\item The uncertainty in the value of $\sin(2\beta)$ has been
      considerably reduced. We find
	\begin{equation}\label{212}
	\sin(2 \beta) = \left\{ \begin{array}{rc}
	0.60 \pm 0.14 & (\rm{range~II}) \\
	0.61 \pm 0.09 & (\rm{range~III})
	\end{array}\right.
	\end{equation}
\item  The uncertainties in $\sin(2\alpha)$ and $\sin(2\gamma)$
         although somewhat reduced remain very large.
\item  For $\mid V_{td} \mid$, $x_s$ and $BR(K^+ \to \pi^+ \nu \bar\nu)$
	we find
	\begin{equation}\label{213}
	\mid V_{td} \mid = \left\{ \begin{array}{rc}
	(9.5 \pm 1.4)\cdot 10^{-3} & (\rm{range~II}) \\
	(9.4 \pm 1.0)\cdot 10^{-3} & (\rm{range~III})
	\end{array}\right.
	\end{equation}
	\begin{equation}\label{214}
	x_s= \left\{ \begin{array}{rc}
	13.3 \pm 4.3 & (\rm{range~II}) \\
	12.9 \pm 2.8 & (\rm{range~III})
	\end{array}\right.
	\end{equation}
	\begin{equation}\label{215}
	BR(K^+ \to \pi^+ \nu \bar\nu) = \left\{ \begin{array}{rc}
	(1.07 \pm 0.24)\cdot 10^{-10} & (\rm{range~II}) \\
	(1.03 \pm 0.15)\cdot 10^{-10} & (\rm{range~III})
	\end{array}\right.
	\end{equation}
\end{itemize}

This exercise implies that if the accuracy of various parameters given
in (\ref{210}) and (\ref{211}) is achieved the determination of
$\mid V_{td} \mid$ and the predictions for $\sin(2\beta)$ and $BR(K^+
\to \pi^+ \nu \bar\nu)$  are quite accurate. A sizable uncertainty in
$x_s$ remains however.

Another important message from this analysis is the inability of a
precise determination of $\sin(2\alpha)$ and $\sin(2\gamma)$ on the
basis of $\varepsilon_{K}$, $B^o - \bar{B^o}$, $|V_{cb}|$ and
$|V_{ub}/V_{cb}|$ alone. Although the great sensitivity of
$\sin(2\alpha)$ and $\sin(2\gamma)$ to various parameters has been
already stressed by several authors, in particular in
refs.~\cite{alilondon:93,dibdunietzgilman:90,gilmannir:90,lusignoli:92},
 our analysis shows that even with
the improved values of the parameters in question as given in
(\ref{210}) and (\ref{211}) a precise determination of $\sin(2\alpha)$
and $\sin(2\gamma)$  should not be expected in this millennium.

The fact that $\sin(2\beta)$ can be much easier determined than
$\sin(2\alpha)$ and $\sin(2\gamma)$ is easy to understand. Since $R_t$
is generally by at least a factor of two larger than $R_b$, the angle
$\beta$ is much less sensitive to the changes in the position of the
point $A = (\bar\varrho,\bar\eta)$ in the unitarity triangle than the
remaining two angles.

\subsection{The Impact of $BR(K^+ \to \pi^+ \nu \bar\nu)$ and $x_d/x_s$}

$BR(K^+ \to \pi^+ \nu \bar\nu)$ and $x_d/x_s$ determine $ \mid V_{td}
\mid$ and $R_t$. If our expectations for the ranges discussed above are
correct we should be able to have a rather accurate prediction for
$BR(K^+ \to \pi^+ \nu \bar\nu)$ using the analysis of $\varepsilon_{K}$
and of $B_d^o - \bar{B_d^o}$ mixing. Measuring $BR(K^+ \to \pi^+ \nu
\bar\nu)$ to similar accuracy would either confirm the standard model
predictions or indicate some physics beyond the standard model.

We infer from tabs.~\ref{tab:rangeb} and \ref{tab:rangec} that measurements
of $BR(K^+ \to \pi^+ \nu
\bar\nu)$ with the accuracy of $\pm 10\%$ would be very useful in this
respect.

The accuracy of predictions for $x_s$ are poorer as seen in
(\ref{214}). A measurement of $x_s$ at a $\pm 10\%$ level will have
therefore a considerable impact on the determination of the CKM
parameters and in particular $R_t$ (see (\ref{107c})) provided
$R_{ds}$ is known within $10\%$ accuracy. A numerical exercise is
presented in subsection~\ref{subsec:mtop}.

\subsection{The Impact of CP-asymmetries in B-decays}
\label{subsec:impact}

Measuring the CP-asymmetries in neutral B-decays will give the
definitive answer whether the CKM description of CP violation is
correct. Assuming that this is in fact the case, we want to
investigate the impact of the measurements of $\sin(2\phi_i)$ on the
determination of the unitarity triangle.

Since in the rescaled triangle of fig.~\ref{fig:triangle}
 one side is known,
it suffices to measure two angles to determine the triangle
completely.

It is well known that the measurement of the CP-asymmetry in the decay
$B^o \to \psi K_s $ should give a measurement of $\sin(2\beta)$ without
any theoretical uncertainties. One expects that prior to LHC
experiments the error on $\sin(2\beta)$ should amount roughly to
$\Delta \sin(2\beta) = \pm 0.06$
\cite{cassel:93,babar:93,albrechtetal:92}.  The measurement of
$\sin(2\alpha)$ is more difficult. It requires in addition the
measurement of several channels in order to eliminate the penguin
contributions. An error $\Delta \sin(2\alpha) = \pm 0.10$ prior to LHC
could however be achieved at a SLAC B-factory \cite{babar:93}.

In fig.~\ref{fig:sinesab} we show the impact of such measurements
and also plot the curve (\ref{113d}) which represents superweak models.
Specifically we take
\begin{equation}\label{220}
\sin(2 \beta) = \left\{ \begin{array}{r}
0.60 \pm 0.18 \qquad (\rm{a}) \\
0.60 \pm 0.06 \qquad (\rm{b})
\end{array}\right.
\end{equation}
as an illustration of two measurements of $\sin(2\beta)$ with two
different accuracies.
 Next we take the following three choices
for $\sin(2\alpha)$
\begin{equation}\label{221}
\sin(2 \alpha) = \left\{ \begin{array}{rc}
-0.20 \pm 0.10 & (\rm{I})  \\
0.10 \pm 0.10  & (\rm{II}) \\
0.70 \pm 0.10  & (\rm{III})
\end{array}\right.
\end{equation}

\fig{
\begin{figure}[htb]
}{
\vspace{0.05in}
\centerline{
% adjust actual size of fig. in text here
\epsfysize=3in
%\epsffile{sin2abtot.ps}
% use this if rotation of fig. is needed
 \rotate[r]{
 \epsffile{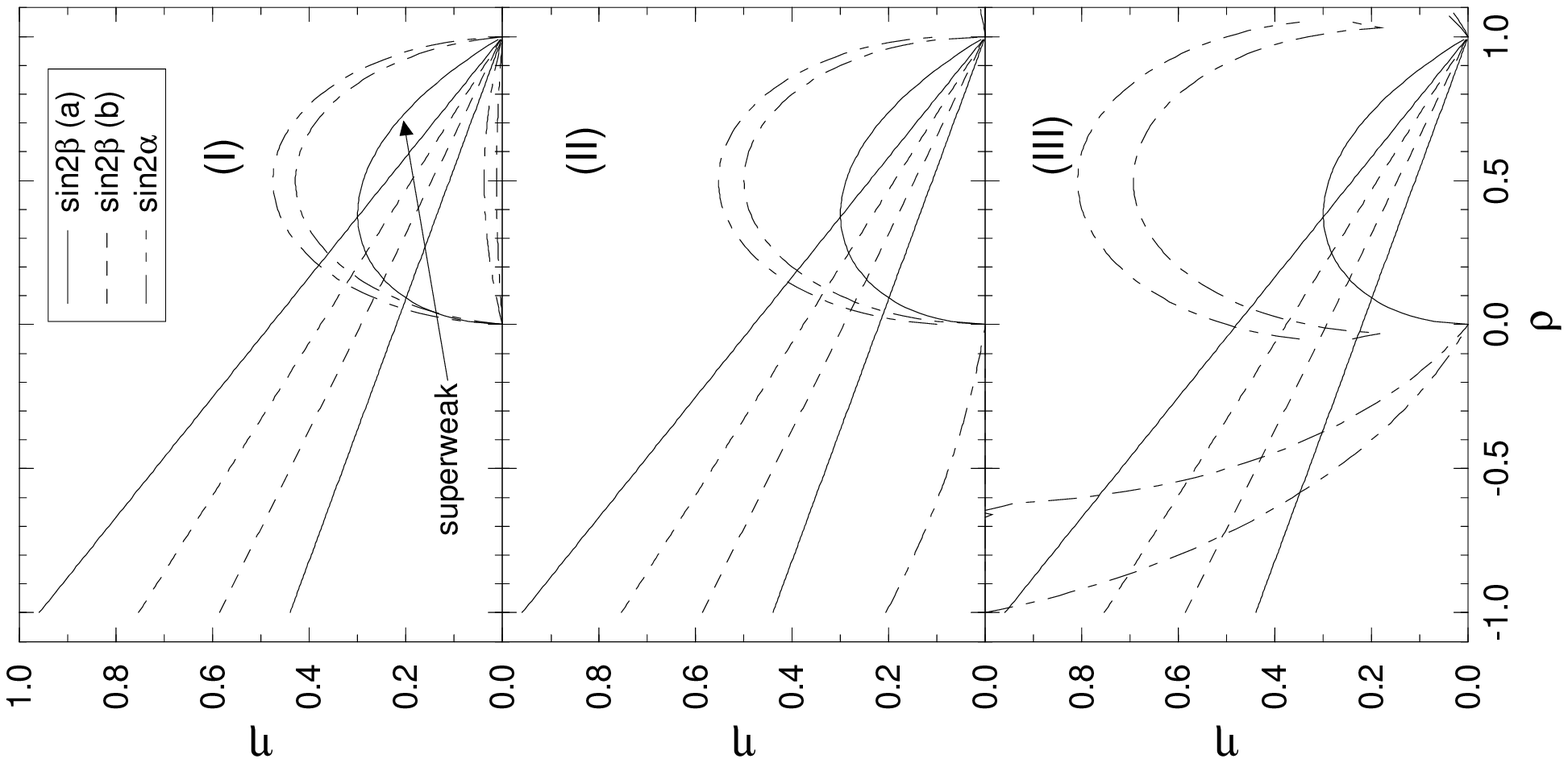}
 }
}
\vspace{0.05in}
}{
\caption[]{\small\sl
	Determination of the unitarity triangle in the
      $(\bar\varrho,\bar\eta)$ plane by measuring $\sin(2\beta)$
	and $\sin(2\alpha)$ as of eqs.~(\ref{220}) and (\ref{221}),
      respectively.
	For $\sin(2\alpha)$ we always find two solutions in $(\bar\varrho,
	\bar\eta)$ and for $\sin(2\beta)$ we only use the solution consistent
	with $\mid V_{ub}/V_{cb} \mid \leq 0.1$.
\label{fig:sinesab}}
\end{figure}
}

In fig.~\ref{fig:sinesbc} we replace the impact of $\sin(2\alpha)$
by the impact of a measurement of $\sin(2\gamma)$ keeping
$\sin(2\beta)$ unchanged. We choose the following values:
\begin{equation}\label{222}
\sin(2 \gamma) = \left\{ \begin{array}{rc}
-0.50 \pm 0.10 & (\rm{I})  \\
    0 \pm 0.10  & (\rm{II}) \\
0.50 \pm 0.10 & (\rm{III})
\end{array}\right.
\end{equation}

\fig{
\begin{figure}[htb]
}{
\vspace{0.05in}
\centerline{
% adjust actual size of fig. in text here
\epsfysize=3in
%\epsffile{sin2bctot.ps}
% use this if rotation of fig. is needed
 \rotate[r]{
 \epsffile{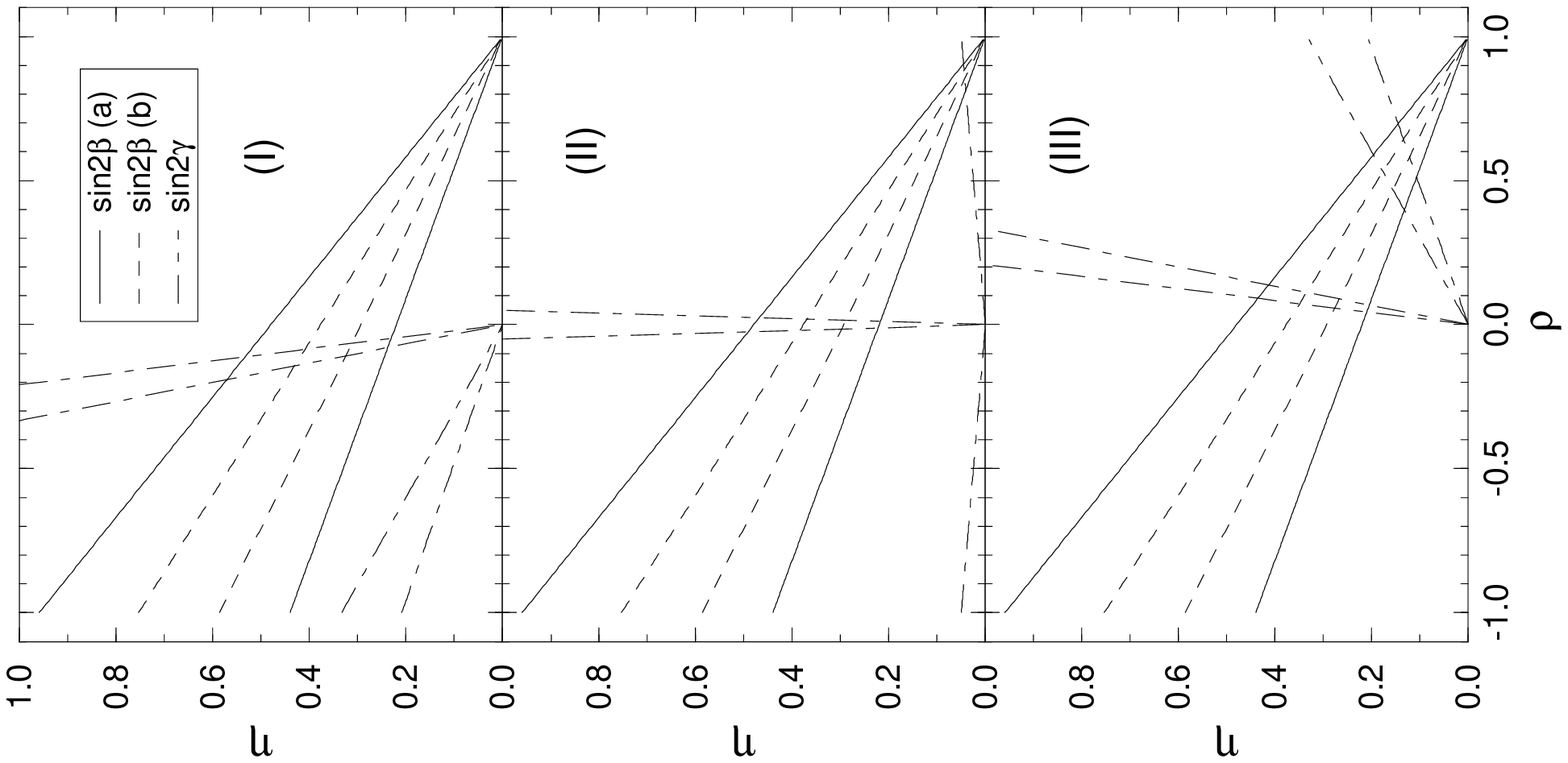}
 }
}
\vspace{0.05in}
}{
\caption[]{\small\sl
      Determination of the unitarity triangle in the
      $(\bar\varrho,\bar\eta)$ plane by measuring $\sin(2\beta)$
      and $\sin(2\gamma)$ as of eqs.~(\ref{220}) and (\ref{222}),
      respectively.
      For $\sin(2\gamma)$ we always find two solutions in $(\bar\varrho,
      \bar\eta)$ and for $\sin(2\beta)$ we only use the solution consistent
      with $\mid V_{ub}/V_{cb} \mid \leq 0.1$.
\label{fig:sinesbc}}
\end{figure}
}

We observe that the measurement of $\sin(2\alpha)$ or $\sin(2\gamma)$
in conjunction with $\sin(2\beta)$ at the expected precision will have
a large impact on the accuracy of the determination of the
unitarity triangle and of the CKM parameters. In order to show this
more explicitly we take as an example:
\begin{equation}\label{223}
\sin(2\beta) = 0.60 \pm 0.06 \qquad \sin(2\alpha) = 0.10 \pm 0.10
\end{equation}
and give in tab.~\ref{ranges2as2b} the predicted ranges for $\delta$,
$\sin(2\gamma)$, $BR(K^+ \to \pi^+ \nu \bar\nu)$, $\mid V_{td} \mid$
and $x_s$ corresponding to the values of $\sin(2\beta)$ and
$\sin(2\alpha)$ given in (\ref{223}) and $ \mid V_{cb} \mid$, $x_d$ and
$m_t$ of (\ref{210}). We use only the solution of $\sin(2\beta)$
consistent with $\mid V_{ub}/V_{cb} \mid \leq 0.1$.

\begin{table}[htb]
\caption[]{
	\small\sl
	Predicted ranges for various quantities calculated by restricting
	$\sin(2\alpha)$ and $\sin(2\beta)$ to the ranges of (\ref{223})
      and using $ \mid V_{cb} \mid$, $x_d$ and $m_t$ of (\ref{210}).
      There is no allowed solution for the second quadrant.
}
\vspace{0.02in}
\begin{center}
\begin{tabular}{|c||c|c|}
\hline
    & Min & Max  \\
\hline
\hline
$\delta$ & 69.5 & 77.8 \\
\hline
$\sin(2 \gamma)$  &   0.42 &   0.66  \\
\hline
$\mid V_{td} \mid\cdot 10^3$  &  8.4 &  9.1  \\
\hline
$x_s$		   &  15.0 &  17.5  \\
\hline
$BR(K^+ \to \pi^+ \nu \bar\nu)  \cdot 10^{10}$
 &   0.90 &   1.12  \\
\hline
\end{tabular}
\end{center}
\label{ranges2as2b}
\end{table}

It should be stressed that this impressive accuracy can only be
achieved by measuring $\sin(2\alpha)$ or $\sin(2\gamma)$ in addition to
$\sin(2\beta)$. This is easy to understand in view of the fact that
the expected accuracy of the measurements of $\sin(2\alpha)$
and $\sin(2\gamma)$ is considerably higher than the corresponding
accuracy of the predictions on basis of $\varepsilon_{K}$, $B^o - \bar{B^o}$
mixing, $\mid V_{ub}/V_{cb} \mid$ and $\mid V_{cb} \mid$ alone.

\subsection{$K^+ \to \pi^+ \nu \bar\nu$,
	$\sin(2\beta)$, $ \mid V_{cb} \mid$, $m_t$ and $x_d/x_s$ }
\label{subsec:mtop}

We would like to address now our last question posed in the
introduction:

How well should one measure $BR(K^+ \to \pi^+ \nu \bar\nu)$,
$\sin(2\beta)$, $ \mid V_{cb} \mid$, $m_t$ and $x_d/x_s$
 in order to obtain an
acceptable determination of the CKM matrix on the basis of these five
quantities alone. As we stated at the beginning of this paper
$K^+ \to \pi^+ \nu \bar\nu$ and $\sin(2\beta)$ are essentially free of
any theoretical uncertainties. $\mid V_{cb} \mid$ on the other hand is
easier to determine than $\mid V_{ub}/V_{cb} \mid$ and once the top
quark is discovered $m_t$ should be known relatively well. Finally
$x_d/x_s$ determines directly $R_t$ by means of eq.~(\ref{107c}).

In fig.~\ref{fig:sin2bbr} we show the result of this exercise taking
(\ref{Lmsmcrange2}) and
\begin{equation}\label{230}
\sin(2\beta)  =  0.60 \pm 0.06 \quad \mid V_{cb} \mid  =  0.040
\pm 0.001 \quad m_t  = (170 \pm 5)~GeV
\end{equation}
\begin{equation}\label{230a}
BR(K^+ \to \pi^+ \nu \bar\nu)   = \left\{ \begin{array}{rc}
(1.0 \pm 0.2) \cdot 10^{-10} & (\rm{I})  \\
(1.0 \pm 0.1) \cdot 10^{-10} & (\rm{II})
\end{array}\right.
\end{equation}

In tab.~\ref{tab:sin2bbr} we give the predicted ranges of various quantities
for the two cases considered.

\begin{table}[htb]
\caption[]{
	 \small\sl
	Ranges of various quantities calculated with constraints from
      eqs.~(\ref{230}) and (\ref{230a}) .
}
\vspace{0.02in}
\begin{center}
\begin{tabular}{|c||c|c||c|c|}
\hline
\multicolumn{1}{|c||}{  } & \multicolumn{2}{c||}{(I)} &
\multicolumn{2}{c|}{(II)} \\
\hline
    & Min & Max & Min & Max \\
\hline
\hline
$\sin(2 \alpha)$        &  -0.917 &   0.978 &   -0.691 &   0.973  \\
\hline
$\sin(2 \gamma)$         &   -0.704 &   1.000 &  -0.418 &   0.976 \\
\hline
$\mid V_{td} \mid\cdot 10^3$    &  6.9 & 10.3 &  7.6 & 9.7  \\
\hline
\end{tabular}
\end{center}
\label{tab:sin2bbr}
\end{table}

In addition  we show in fig.~\ref{fig:sin2bbr} the result of a possible
measurement of $x_d/x_s$ corresponding to $R_t = 1.0 \pm 0.1$. We
observe that provided the expected accuracy of measurements is
achieved we should have a respectable determination of $\mid V_{td}
\mid$ this way. Fig.~\ref{fig:sin2bbr} indicates that for the
$\Delta V_{cb}$ and $\Delta m_t$ assumed here, $BR(K^+ \to \pi^+ \nu
\bar\nu)$ must be measured with a precision of $\pm 10\%$ to be
competitive with $\Delta R_t = \pm 10\%$ extracted hopefully one day
from $x_d/x_s$. The uncertainty in the predictions for $\sin(2\alpha)$
and $\sin(2\gamma)$ is very large as in the analysis of
subsection~\ref{subsec:look}.

\fig{
\begin{figure}[htb]
}{
\vspace{0.05in}
\centerline{
% adjust actual size of fig. in text here
\epsfysize=5in
%\epsffile{sin2abBR.ps}
% use this if rotation of fig. is needed
 \rotate[r]{
 \epsffile{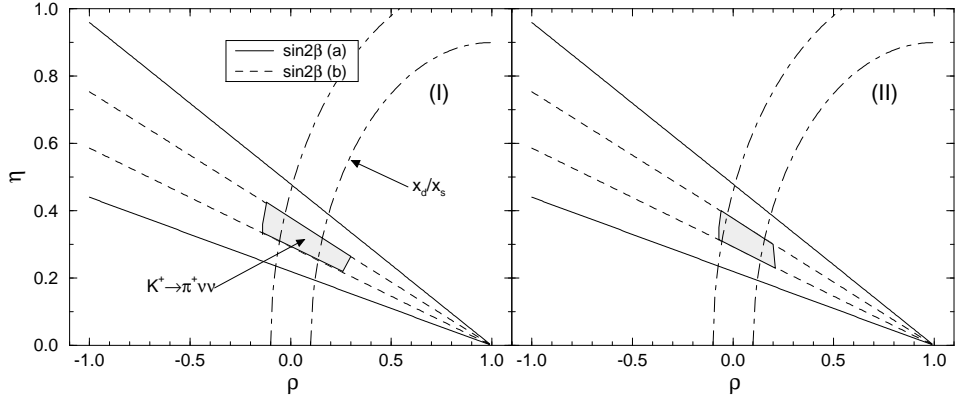}
 }
}
\vspace{0.05in}
}{
\caption[]{\small\sl
   Allowed ranges in the $(\bar\varrho, \bar\eta)$ plane with constraints
   from eqs.~(\ref{230}) and (\ref{230a}) for $BR(K^+ \to \pi^+ \nu \bar\nu)$
   and $R_t = 1.0 \pm 0.1$.
}
\label{fig:sin2bbr}
\end{figure}
}

\subsection{$\varepsilon_K$, $B_d^o - \bar{B}_d^o$ Mixing, $\sin(2\beta)$
and $\sin(2\alpha)$}

It is useful to combine the results of subsections \ref{subsec:first},
\ref{subsec:look} and \ref{subsec:impact} by making the customary
$\sin(2\beta)$ versus $\sin(2\alpha)$ plot \cite{nir:74}. This plot
demonstrates very clearly the correlation between $\sin(2\alpha)$ and
$\sin(2\beta)$. The allowed ranges for $\sin(2\alpha)$ and
$\sin(2\beta)$ corresponding to the choices of the parameters in
(\ref{200}), (\ref{210}) and (\ref{211}) are shown in
fig.~\ref{fig:sin2bvs2a} together with the results of the independent
measurements of $\sin(2\beta) = 0.60 \pm 0.06$ and $\sin(2\alpha)$
given by (\ref{221}). The latter are represented by dark shaded rectangles.
The black rectangles illustrate the accuracy of future LHC measurements
($\Delta\sin(2\alpha) = \pm 0.04$, $\Delta\sin(2\beta) = \pm 0.02$)
\cite{camilleri:93}.

\fig{
\begin{figure}[htb]
}{
\vspace{0.05in}
\centerline{
% adjust actual size of fig. in text here
\epsfysize=5in
%\epsffile{sin2bvs2a.ps}
% use this if rotation of fig. is needed
 \rotate[r]{
 \epsffile{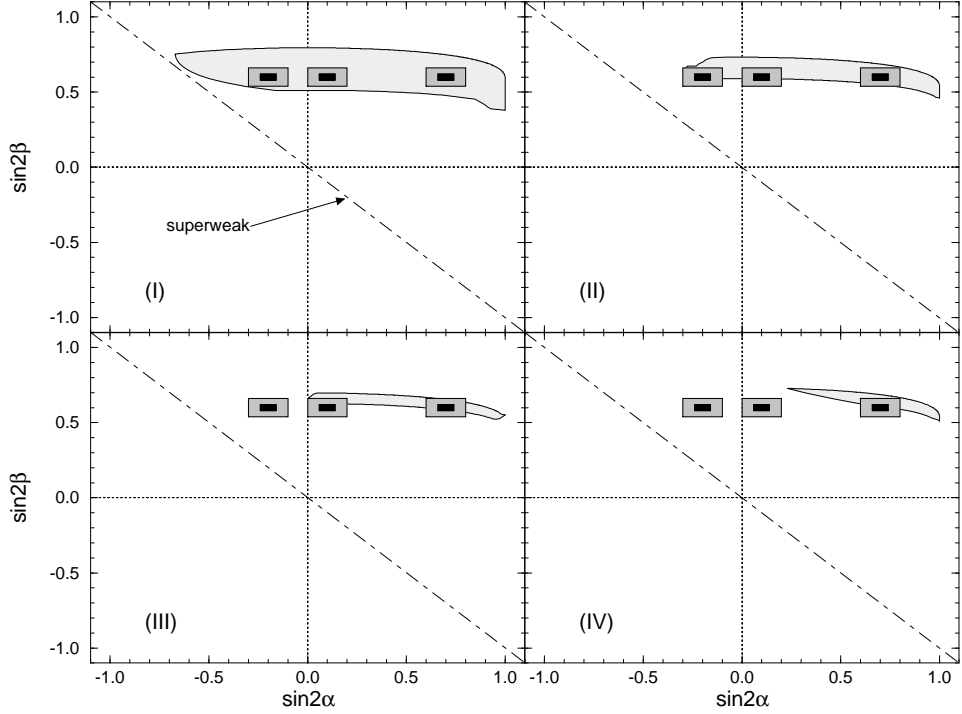}
 }
}
\vspace{0.05in}
}{
\caption[]{\small\sl
$\sin(2\alpha)$ versus $\sin(2\beta)$ plot corresponding to the parameter
   ranges I-IV as of (\ref{200}), (\ref{210}), (\ref{211}) and (\ref{231})
   and the dark shaded rectangles given by (\ref{221}) and (\ref{220})\,(b).
   The black rectangles illustrate the accuracy of future LHC measurements.
\label{fig:sin2bvs2a}}
\end{figure}
}

We also show the results of an analysis in which the
accuracy of various parameters is as in (\ref{210}) but with the central
values modified:

\bigskip
\underline{Range IV}

\begin{equation}\label{231}
\begin{array}{rclrcl}

\left| V_{cb} \right| & = &  0.038 \pm 0.002 &
\mid V_{ub}/V_{cb} \mid & = & 0.08 \pm 0.01 \\
B_K & = & 0.70 \pm 0.07 & \sqrt{B_{B_d}} F_{B_d} & = & (185 \pm 15)~MeV \\
x_d & = & 0.72 \pm 0.04 & m_t & = & (165 \pm 7)~GeV \\
\end{array}
\end{equation}

In addition we show the prediction of superweak theories which in this plot
is represented by a straight line.

There are several interesting features on this plot:

\begin{itemize}
\item The impact of the direct measurements of $\sin(2\beta)$ and
	$\sin(2\alpha)$ is clearly visible in this plot
\item In cases III and IV we have examples where the measurements of
	$\sin(2\alpha)$ are incompatible with the predictions coming
	from $\varepsilon_{K}$ and $B^o - \bar{B^o}$ mixing. This would be a
	signal for physics beyond the standard model. The measurement
	of $\sin(2\alpha)$ is essential for this.
\item The case IV shows that for a special choice of parameters the
	predictions for the asymmetries coming from $\varepsilon_{K}$, $B^o -
	\bar{B^o}$ mixing, $\mid V_{cb} \mid$ and $\mid V_{ub}/V_{cb} \mid$
	can be quite accurate when these four constraints can only be
	satisfied simultaneously in a small area of the $(\bar\varrho,
	\bar\eta)$~space. Decreasing $\mid V_{cb} \mid$, $\mid
	V_{ub}/V_{cb} \mid$ and $m_t$ and increasing $F_B$ would make the
	allowed region in the case IV even smaller.
\item We also observe that the future measurements of asymmetries and
	the improved ranges for the parameters relevant for $\varepsilon_{K}$
	and $B^o - \bar{B^o}$ mixing will probably allow to rule out the
	superweak models.
\end{itemize}

\newsection{Summary and Conclusions}

The top quark discovery and the measurements of $BR(K^+ \to \pi^+ \nu
\bar\nu)$, $x_s$ and of CP violating asymmetries in B-decays will play
crucial roles in the determination of the CKM parameters and in the
tests of the standard model. Similarly the improvements in the
determination of the CKM elements $V_{ub}$ and $V_{cb}$
in tree  level B-decays and the improved calculations of the
non-perturbative parameters like $B_K$ and $\sqrt{B_B} F_B$ will
advance our understanding of weak decay phenomenology. In this paper
we have made an excursion in the future trying to see what one could
expect in this field in the coming five to ten years prior to LHC
experiments.

In the first part of the numerical analysis we have investigated how
the top quark discovery together with the improved determinations of
$\mid~V_{ub}/V_{cb}~\mid$, $\mid~V_{cb}~\mid$, $B_K$ and $\sqrt{B_B}
F_B$ would allow for the determination of the unitarity triangle and
more accurate predictions for $K^+ \to \pi^+ \nu \bar\nu$, $B_s^o -
\bar{B_s^o}$~mixing and $\sin(2\phi_i)$. Our main findings in this
part can be summarized as follows:
\begin{itemize}
\item We expect that around  the year 2000 satisfactory predictions for
	 $\mid V_{td} \mid$, $\sin(2\beta)$ and $BR(K^+ \to \pi^+ \nu
	\bar\nu)$ should be possible.
\item   A sizeable uncertainty in $x_s$ and huge uncertainties in
	$\sin(2\alpha)$ and in $\sin(2\gamma)$ will remain however.
\end{itemize}

In the second part of our analysis we have investigated the impact of
future measurements of $BR(K^+ \to \pi^+ \nu \bar\nu)$, $x_s$ and
$\sin(2\phi_i)$. Our main findings in this second part can be
summarized as follows:
\begin{itemize}
\item The measurements of $\sin(2\alpha)$, $\sin(2\beta)$ and
	$\sin(2\gamma)$ will have an impressive impact on the determination of
	the CKM parameters and the tests of the standard model.
\item This impact is further strengthened by combining the constraints
	considered in the two parts of our analysis as seen most clearly in
	fig.~\ref{fig:sin2bvs2a}.
\item Future LHC B-physics experiments around the year 2005 will refine
      these studies as evident from fig.~\ref{fig:sin2bvs2a} and
      ref.~\cite{camilleri:93}.
\end{itemize}

In our analysis we have concentrated on quantities which have either
been already measured $(\varepsilon_K, x_d)$ or quantities which are
practically free from theoretical uncertainties such as $x_d/x_s$,
$K^+ \to \pi^+ \nu \bar\nu$ and certain asymmetries in B-decays. We
however stress at this point that the measurements of
$\varepsilon^{\prime} / \varepsilon$, $B \to s \gamma$, $K_L \to \mu^+
\mu^{-}$, $K_L \to \pi^o e^+ e^-$, $K_L \to \pi^o \nu \bar\nu$
and other rare decays discussed in the literature are also
very important for our understanding of weak decays. In particular a
measurement of a non-zero $Re(\varepsilon^{\prime} / \varepsilon)$
to be expected in few years from now, will give most probably first
signal of direct CP violation. Unfortunately, all these decays are
either theoretically less clean than the decays  considered here or
they are more difficult to measure. Clearly some dramatic improvements
in the experimental techniques and in non-perturbative methods  could
change this picture in the future.

We hope that our investigations and the analytic formulae derived in
this paper will facilitate the waiting for $m_t$, $K^+ \to \pi^+ \nu
\bar\nu$, $B_s^o - \bar{B_s^o}$~mixing and CP asymmetries in B-decays.
There is clearly a very exciting time ahead of us.

%%%%%%%%%%%%%%%%%%%%%%%%%%%%%%%%%%%%%%%%%%%%%%%%%%%%%%%%%%%%%%%%%%%%%%%%%
% Acknowledgment and bibliography
%%%%%%%%%%%%%%%%%%%%%%%%%%%%%%%%%%%%%%%%%%%%%%%%%%%%%%%%%%%%%%%%%%%%%%%%%

\vskip 1cm
\begin{center}
{\large\bf Acknowledgement}
\end{center}
\noindent
A.J.~Buras would like to thank the members of the CP-B panel at the
Max-Planck-Institut in Munich for exciting discussions.

\end{document}